\documentclass[usenames,onecolumn,dvipsnames]{aastex63}

\usepackage{graphicx}	
\usepackage{amsmath}	
\usepackage{amssymb}	
\usepackage{relsize}
\usepackage{mathtools}
\usepackage{bigints}
\usepackage{sidecap}
\usepackage{caption,subcaption}

\DeclareMathOperator{\sech}{sech}

\DeclareMathOperator{\erf}{erf}

\usepackage{floatrow}
\usepackage{ulem}
\usepackage{soul}
\usepackage{comment}
\usepackage{bm}

\usepackage{macros_ub}
\graphicspath{{./}{figures/}}

\received{XXX}
\revised{YYY}
\accepted{ZZZ}

\begin{document}
\vspace{1mm}


\shorttitle{Phase-space spirals}
\shortauthors{}

\title{A Comprehensive Perturbative Formalism for Phase-Mixing in Perturbed Disks. I. Phase spirals in an Infinite, Isothermal Slab}

\correspondingauthor{Uddipan Banik}
\email{uddipan.banik@yale.edu}

\author[0000-0002-9059-381X]{Uddipan Banik}
\affiliation{Department of Astronomy, Yale University, PO. Box 208101, New Haven, CT 06520, USA}

\author[0000-0003-2660-2889]{Martin~D.~Weinberg}
\affiliation{Department of Astronomy, University of Massachusetts at Amherst, 710 N. Pleasant St., Amherst, MA 01003}

\author[0000-0003-3236-2068]{Frank~C.~van den Bosch}
\affiliation{Department of Astronomy, Yale University, PO. Box 208101, New Haven, CT 06520, USA}

\label{firstpage}

\begin{abstract}
Galactic disks are highly responsive systems that often undergo external perturbations and subsequent collisionless equilibration, predominantly via phase-mixing. We use linear perturbation theory to study the response of infinite isothermal slab analogues of disks to perturbations with diverse spatio-temporal characteristics. Without self-gravity of the response, the dominant Fourier modes that get excited in a disk are the bending and breathing modes, which, due to vertical phase-mixing, trigger local phase-space spirals that are one- and two-armed, respectively. We demonstrate how the lateral streaming motion of slab stars causes phase spirals to damp out over time. The ratio of the perturbation timescale ($\tau_\rmP$) to the local, vertical oscillation time ($\tau_z$) ultimately decides which of the two modes is excited. Faster, more impulsive ($\tau_\rmP < \tau_z$) and slower, more adiabatic ($\tau_\rmP > \tau_z$) perturbations excite stronger breathing and bending modes, respectively, although the response to very slow perturbations is exponentially suppressed. For encounters with satellite galaxies, this translates to more distant and more perpendicular encounters triggering stronger bending modes. We compute the direct response of the Milky Way disk to several of its satellite galaxies, and find that recent encounters with all of them excite bending modes in the Solar neighborhood. The encounter with Sagittarius triggers a response that is at least $1-2$ orders of magnitude larger than that due to any other satellite, including the Large Magellanic Cloud. We briefly discuss how ignoring the presence of a dark matter halo and the self-gravity of the response might impact our conclusions.
\end{abstract}


\section{Introduction}
\label{sec:intro}

The relaxation or equilibration of self-gravitating systems is a ubiquitous astrophysical phenomenon that drives the formation and evolution of star-clusters, galaxies and cold dark matter halos. In quasi-equilibrium, the phase-space density of such collisionless systems can be well characterized by a distribution function (DF) which, according to the strong Jeans theorem, is a function of the conserved quantities or actions of the system. When such a system is perturbed out of equilibrium by a time-dependent gravitational perturbation, either external (e.g., encounter with another galaxy) or internal (e.g., bars or spiral arms), the original actions of the stars are modified, and the system has to re-establish a new (quasi-)equilibrium. Since disk galaxies are highly ordered, low-entropy (i.e., cold) systems, they are extremely responsive. Even small gravitational perturbations can induce oscillations in the disk, which manifest as either standing or propagating waves \citep[see][for a detailed review]{Sellwood.13}. Such oscillations consist of an initially coherent response of stars to a gravitational perturbation. This coherent response is called {\it collective} if its self-gravity is included. Over time, though, the coherence {\it dissipates}, which manifests as relaxation or equilibration and drives the system towards a new quasi-equilibrium, free of large scale oscillations. Equilibration in galactic disks is dominated by collisionless effects, including purely kinematic processes like phase-mixing (loss of coherence in the response due to different orbital frequencies of stars), and self-gravitating or collective processes like Landau damping \citep[loss of coherence due to non-dissipative damping of waves by wave-particle interactions,][]{LyndenBell.62} and violent relaxation \citep[loss of coherence due to scrambling of orbital energies in a time-varying potential,][]{LyndenBell.67}. It is noteworthy to point out that without phase-mixing neither Landau damping \citep[][]{Maoz.91} nor violent relaxation \citep[see][]{Sridhar.89} would result in equilibration. A final equilibration mechanism is chaotic mixing, the loss of coherence resulting from the exponential divergence of neighboring stars on chaotic orbits \citep[e.g.,][]{Merritt.Valluri.96,Daniel.Wyse.15,Banik.vdBosch.22}. As long as most of the phase-space is foliated with regular orbits (i.e., the Hamiltonian is near-integrable), chaotic mixing should not make a significant contribution, and phase-mixing may thus be considered the dominant equilibration mechanism.

Disk galaxies typically reveal out-of-equilibrium features due to incomplete equilibration. These may appear in the form of bars and spiral arms, which are large-scale perturbations in the radial and azimuthal directions, responsible for a slow, secular evolution of the disk. In the vertical direction, disks often reveal warps \citep[][]{Binney.92}. In the case of the Milky Way (hereafter MW) disk, which can be studied in much greater detail than any other system, recent data from astrometric and radial-velocity surveys such as SEGUE \citep[][]{Yanny.etal.09}, RAVE \citep[][]{Steinmetz.etal.06}, GALAH \citep[][]{Bland-Hawthorn.etal.19}, LAMOST \citep[][]{Cui.etal.12} and above all Gaia \citep[][]{Gaia_collab.16, Gaia_collab.18a, Gaia_collab.18b} has revealed a variety of additional vertical distortions. At large galacto-centric radii ($>10 \kpc$) this includes, among others, oscillations and corrugations \citep[][]{Xu.etal.15,Schonrich.Dehnen.18}, and streams of stars kicked up from the disk that undergo phase-mixing, sometimes referred to as `feathers' \citep[e.g.,][]{Price-Whelan.etal.15, Thomas.etal.19, Laporte.etal.22}. Similar oscillations and vertical asymmetries have also been reported in the Solar vicinity \citep[e.g.,][]{Widrow.etal.12, Williams.etal.13, Yanny.Gardner.13, Quillen.etal.18, Gaia_collab.18b, Bennett.Bovy.19, Carrillo.etal.19}. One of the most intriguing structures is the phase-space spiral discovered by \cite{Antoja.etal.18}, and studied in more detail in subsequent studies \citep[e.g.,][]{Bland-Hawthorn.etal.19,Li.Widrow.21,Li.21,Gandhi.etal.22}. Using data from Gaia DR2 \citep[][]{Gaia_collab.18a}, \cite{Antoja.etal.18} selected $\sim 900$k stars within a narrow range of galacto-centric radius and azimuthal angle centered around the Sun. When plotting the density of stars in the $(z,v_z)$-plane of vertical position, $z$, and vertical velocity, $v_z$, they noticed a faint, unexpected spiral pattern, which became more enhanced when colour-coding the $(z,v_z)$-`pixels' by the median radial or azimuthal velocities. The one-armed spiral makes 2-3 complete wraps, resembling a snail shell, and is interpreted as a signature of phase-mixing in the vertical direction following a perturbation, which \cite{Antoja.etal.18} estimate to have occurred between 300 and 900 Myr ago. More careful analyses in later studies \citep[e.g.,][etc.]{Bland-Hawthorn.etal.19,Li.21} have nailed down the interaction time to $\sim 500\Myr$ ago.

The discovery of all these oscillations in the MW disk has ushered in a new, emerging field of astrophysics, known as galactoseismology \citep[][]{Widrow.etal.12, Johnston.etal.17}. Similar to how the timbre of musical notes reveals characteristics of the instrument that produced the sound, the `ringing' of a galactic disk can (in principle) reveal its structure (both stellar disk plus dark matter halo). And similar to how the timbre can tell us whether the string of a violin was plucked (pizzicato) or bowed (arco), the ringing of a galactic disk can reveal information about the perturbation that set the disk ringing. Phase spirals are especially promising in this regard: their structure holds information about the gravitational potential in the vertical direction \citep[in particular, the vertical frequency as a function of the vertical action,][]{Antoja.etal.18} and about the type of perturbation that triggered the phase spiral \citep[e.g., bending mode vs. breathing mode, see][and Section~\ref{sec:impulsive_kick} below]{Widrow.etal.14, Darling.Widrow.19a}. In addition, by unwinding the phase spiral one can in principle determine how long ago the vertical oscillations were triggered. By studying phase spirals at multiple locations in the disk, one may even hope to use some form of triangulation to infer the direction or location from which the perturbation emerged (assuming, of course, that the phase spirals at different locations were all triggered by the same perturbation).

However promising galactoseismology may seem, many questions remain: what kind of perturbation can trigger a phase spiral? how long do phase spirals remain detectable, and what equilibration mechanism(s) causes their demise? Can we really constrain the vertical potential of the disk, or does self-gravity of the perturbation make it difficult to achieve?  What kind of constraints can we infer regarding the perturber that triggered the phase spiral? Is galactoseismology likely to be confusion limited, i.e., should we expect that each location in the disk experiences oscillations arising from multiple, independent perturbations? If so, how does this impact our ability to extract useful information? Answering these questions necessitates a deep understanding of how the MW disk, and disk galaxies in general, respond to perturbations.

To date, these questions have mainly been addressed using numerical $N$-body simulations or fairly simplified analytical approaches. In particular, numerous studies have investigated how the MW disk responds to interactions with the Sagittarius (Sgr) dwarf galaxy \citep[e.g.,][]{Gomez.etal.13, Donghia.etal.16, Laporte.etal.18, Khanna.etal.19,Hunt.etal.21}. While simulations likes these have demonstrated that the interaction with Sgr can indeed spawn phase spirals in the Solar vicinity \citep[][]{Antoja.etal.18,Binney.Schonrich.18, Darling.Widrow.19b, Laporte.etal.19, Bland-Hawthorn.etal.19, Hunt.etal.21, Bennett.etal.21}, none of them have been able to produce phase spirals that match those observed in the Gaia data. As discussed in detail in \cite{Bennett.etal.21} and \cite{Bennett.Bovy.21}, this seems to suggest that the amplitude and shape of the ``Gaia snail" cannot be produced by Sgr alone. An alternative explanation, explored by \citet{Khoperskov.etal.19}, is that the Gaia snail was created by buckling of the MW's bar. However, this explanation faces its own challenges \citep[see e.g.,][]{Laporte.etal.19, Bennett.Bovy.21}. Triggering the Gaia snail with a spiral arm \citep[][]{Faure.etal.14} is also problematic, in that it requires the spiral arms to have unusually large amplitude \citep[][]{Quillen.etal.18}. Clearly then, despite a large number of studies, pinpointing the origin of the phase spiral in the Solar vicinity still remains an unsolved problem.
 
Although simulations have the obvious advantage that they can probe the complicated response of a perturbed disk to a realistic perturbation, which often is analytically intractable, especially if the response is large (non-linear), there are also clear disadvantages. Foremost, reaching sufficient resolution to resolve the kind of fine-structure that we can observe with data sets like Gaia requires extremely large simulations with $N > 10^8-10^9$ particles \citep[][]{Weinberg.Katz.07a, Binney.Schonrich.18, Hunt.etal.21}. Although such simulations are no longer beyond our reach \citep[see e.g.,][]{Bedorf.etal.14, Fujii.etal.19, Hunt.etal.21, Peterson.etal.22}, it is clear that using such simulations to explore large areas of parameter space remains a formidable challenge. To overcome this problem, a semi-analytical approach called the {\it backward-integrating restricted N-body method} was developed originally in the context of perturbation by bars \citep[e.g.,][]{Leeuwin.etal.93,Vauterin.Dejonghe.97,Dehnen.00}, and later on used by \cite{Hunt.Bovy.18} and \cite{Hunt.etal.19} to study non-equilibrium features in the MW caused by transient spiral arms. This method is effectively a Lagrangian formalism to solve the collisionless Boltzmann equation (hereafter CBE) by integrating test particles in the perturbed potential in a restricted N-body framework, i.e., without self-consistently developing the potential perturbation from the DF perturbation. Although appropriate for studying the local kinematic distribution of particles, this approach becomes too expensive to study the global equilibration of a system. Hence, it is important to consider alternative analytical methods that can be used to investigate the global response of a disk.

In this vein, this paper presents a rigorous, perturbative, Eulerian formalism to compute the response of a disk to perturbations. In order to gain valuable insight into the physical mechanism of phase-mixing, without resorting to the computational complexity involved in modelling a realistic disk, which we postpone to Paper~II (Banik et al., in preparation), in this first paper in the series we consider perturbations of an infinite slab with a vertical profile, but homogeneous in the lateral directions. Although a poor representation of a realistic galactic disk, this treatment captures most of the essential features of how disks respond to gravitational perturbations. We study the response of the slab to perturbers of various spatial and temporal scales, with a focus on the formation and dissolution of phase spirals resulting from the vertical oscillations and phase-mixing of stars.

This paper is organized as follows. Section~\ref{sec:linear_theory} describes the application of perturbation theory to our infinite, isothermal slab. Section~\ref{sec:impulsive_kick} then uses these results to work out the response to an impulsive, single-mode perturbation, which nicely illustrates how phase spirals originate from vertical oscillations and how they damp out due to lateral mixing. Sections~\ref{sec:localized} and~\ref{sec:non-impulsive} generalize this to responses to localized (wave packet) and non-impulsive perturbations, respectively. In Section~\ref{sec:sat_encounter} we investigate the response to satellite encounters and examine which satellite galaxies in the halo of the MW can trigger bending and/or breathing modes strong enough to trigger phase spirals at the Solar radius (still approximating the MW disk as an infinite, isothermal slab). We summarize our findings in Section~\ref{sec:concl}.

\newpage
\section{Linear perturbation theory for collisionless systems}
\label{sec:linear_theory}

\subsection{Linear perturbative formalism}

Let the unperturbed steady state distribution function (DF) of a collisionless stellar system be given by $f_0$ and the corresponding Hamiltonian be $H_0$. $f_0$ satisfies the CBE,
\begin{align}
[f_0,H_0]=0,
\end{align}
where the square brackets correspond to the Poisson bracket. Now let us introduce a small time-dependent perturbation in the potential, $\Phi_\rmP(t)$, such that the perturbed Hamiltonian becomes
\begin{align}
H=H_0+\Phi_\rmP(t)+\Phi_1(t),
\end{align}
where $\Phi_1$ is the gravitational potential sourced by the response density, $\rho_1 = \int f_1 d^3\bv$, via the Poisson equation,
\begin{align}
\nabla^2\Phi_1=4\pi G\rho_1.
\end{align}
Here $f_1$ is the linear order perturbation in the DF, i.e., the linear response of the system to the perturbation in the potential. The perturbed DF can thus be written as
\begin{align}
f=f_0+f_1.
\end{align}
Assuming that the perturbations are small such that linear perturbation theory holds, the time-evolution of $f_1$ is governed by the following linearized version of the CBE
\begin{align}
\frac{\partial f_1}{\partial t}+[f_1,H_0]+[f_0,\Phi_\rmP]+[f_0,\Phi_1]=0.
\label{CBE_perturb}
\end{align}
In this paper we shall neglect the self-gravity of the disk, i.e., neglect the polarization term, $[f_0,\Phi_1]$, in the lhs of the linearized CBE. We briefly discuss the impact of self-gravity in Section~\ref{sec::caveats}, leaving a more detailed analysis including self-gravity to a forthcoming publication.

\subsection{Hybrid perturbative formalism for an infinite slab}
\label{sec:slab}

We consider the simplified case of perturbations in an infinitely extended slab, uniform in $(x,y)$, but characterized by a vertical density profile $\rho(z)$. Although a rather poor approximation of a realistic galactic disk, this idealized case serves to highlight some of the main characteristics of disk response. We consider perturbations that can be described by a profile in the vertical $z$-direction and by a superposition of plane waves along the $x$-direction, such that $\Phi_\rmP$ and $f_1$ are both independent of $y$. After making a canonical transformation from the phase-space variables $(z,v_z)$ to the corresponding action angle variables $(I_z,w_z)$, Equation~(\ref{CBE_perturb}) becomes
\begin{align}
\frac{\partial f_1}{\partial t}+\frac{\partial H_0}{\partial I_z}\frac{\partial f_1}{\partial w_z}+\frac{\partial H_0}{\partial v_x}\frac{\partial f_1}{\partial x}-\frac{\partial \Phi_\rmP}{\partial w_z}\frac{\partial f_0}{\partial I_z}-\frac{\partial \Phi_\rmP}{\partial x}\frac{\partial f_0}{\partial v_x}=0.
\label{CBE_perturb1}
\end{align}
The unperturbed Hamiltonian $H_0$ can be written as
\begin{align}
H_0 = \frac{v^2_x+v^2_y}{2} + \frac{v^2_z}{2} + \Phi_z(z),
\end{align}
where $v_x$, $v_y$ and $v_z$ are the unperturbed velocities of stars along $x$, $y$ and $z$ respectively, and $\Phi_z(z)$ is the unperturbed potential that dictates the oscillatory vertical motion of the stars.
We expand $\Phi_\rmP$ and $f_1$ as Fourier series that are discrete along $z$ but continuous along $x$:
\begin{align}
\Phi_\rmP(z,x,t)&=\sum_{n=-\infty}^{\infty}\int d k\, \exp{\left[i (n w_z + k x)\right]}\, \Phi_{nk}(I_z,t),\nonumber \\
f_1(z,v_z,x,v_x,v_y,t)&=\sum_{n=-\infty}^{\infty}\int d k\, \exp{\left[i (n w_z + k x)\right]}\, f_{1nk}(I_z,v_x,v_y,t).
\end{align}
Here $z$ can be expressed as the following implicit function of $w_z$ and $I_z$,
\begin{align}
w_z = \Omega_z \int_0^{z} \frac{d z'}{\sqrt{2\left[E_z(I_z)-\Phi_z(z')\right]}}.
\label{z_wz_Iz}
\end{align}
where $\Omega_z=\Omega_z(I_z)$ is the vertical frequency of stars with vertical action $I_z$, given in equation~(\ref{omzvx}) below.

Here and throughout this paper we express any dependence on the continuous wave number $k$ with an index rather than an argument, i.e., $\Phi_{nk}(I_z,t)$ rather than $\Phi_n(k,I_z,t)$. This implies that any function that carries $k$ as an index is in Fourier space.

\begin{figure*}[h!]
  \centering
  \includegraphics[width=0.95\textwidth]{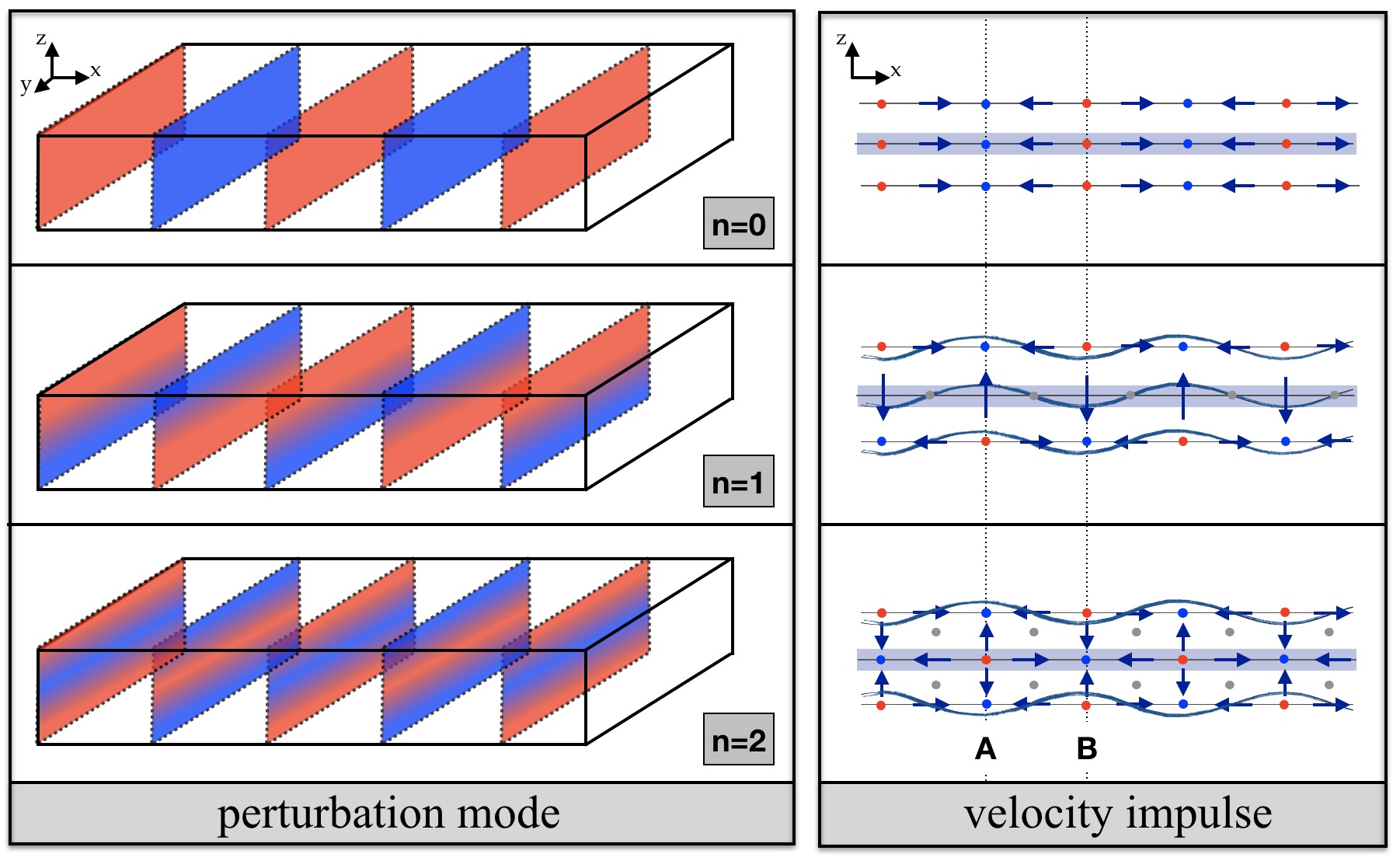}
  \caption{Illustration of the $n=0$, $n=1$ and $n=2$ plane-wave perturbation modes in a laterally uniform and vertically isothermal slab (left-hand panel) and the velocity impulses corresponding to these modes (right-hand panel) in the case of an instantaneous/impulsive perturbation. In the left-hand panel, the rectangular box indicates a random section of the slab, centered on the slab's midplane ($z=0$), while red and blue colors indicate positive and negative $\Phi_\rmP$. For clarity, this color coding is only shown at the extrema (peaks and troughs) of the mode, which has a wave-vector that is pointing in the $x$-direction. The right-hand panel shows an edge-on view of the slab, with arrows indicating the local direction of the velocity impulse caused by the instantaneous perturbation $\Phi_\rmP$, and dots marking locations in the disk where the velocity impulse is zero. Whereas the $n=0$ mode corresponds to a longitudinal perturbation, both $n=1$ and $n=2$ correspond to transverse perturbations; the former is a bending mode, while the latter is a breathing mode (note though that both these modes also cause velocity impulses in the lateral directions). Finally, `A' and `B' mark two specific locations in the slab to which we refer in the text and in Figs.~\ref{fig:impulsive_slab_n1} and~\ref{fig:impulsive_slab_n2}.}
  \label{fig:modes}
\end{figure*}

We express the perturber potential and the DF perturbation or response as linear superpositions of Fourier modes. Since we do not take into account the self-gravity of the response itself, i.e., do not self-consistently solve the Poisson equation along with the CBE, these are not dynamical or normal modes of the system. In other words, the oscillation frequencies of the Fourier modes are just the unperturbed frequencies, $\Omega_z$, and do not follow a dispersion relation as in the self-gravitating case. To aid the visualization of the various Fourier modes, Fig.~\ref{fig:modes} illustrates what the $n=0$, $n=1$ and $n=2$ modes for one particular value of the wavenumber $k$ look like. The figure also indicates the direction of the velocity impulses resulting from an instantaneous perturbation of each mode.

Substitution of the above expressions in equation~(\ref{CBE_perturb1}) yields the following evolution equation for $f_{1nk}$
\begin{align}
\frac{\partial f_{1nk}}{\partial t}+i(n\Omega_z+k v_x)f_{1nk}=i\left(n\frac{\partial f_0}{\partial I_z}+k\frac{\partial f_0}{\partial v_x}\right)\Phi_{nk},
\label{f1nk_de}
\end{align}
where we have used that
\begin{align}
\Omega_z = \frac{\partial H_0}{\partial I_z}\,, \;\;\; v_x=\frac{\partial H_0}{\partial v_x}.
\label{omzvx}
\end{align}
The above first order differential equation in time is easily solved using the Green's function technique. With the initial condition, $f_{1nk}(t_\rmi)=0$, we obtain the following integral form for $f_{1nk}$ for a given time dependence of the perturber potential,
\begin{align}
f_{1nk}(I_z,v_x,v_y,t)&=i\left(n\frac{\partial f_0}{\partial I_z}+k\frac{\partial f_0}{\partial v_x}\right)\int_{t_i}^{t}d\tau \exp{\left[-i(n\Omega_z+k v_x)(t-\tau)\right]}\, \Phi_{nk}(I_z,\tau).
\label{f1nk_slabsol}
\end{align}
This solution is analogous to the particular solution for a forced oscillator with natural frequencies, $n\Omega_z$ and $k v_x$, which is being forced by an external perturber potential, $\Phi_{nk}$. The time-dependence of this external perturbation ultimately dictates the temporal evolution of the perturbation in the DF, $f_{1nk}$. A {\it net} response requires gradients in the (unperturbed) DF with respect to the actions and/or velocities. Similar solutions for the response of perturbed, collisionless systems have been derived in a number of previous studies \citep[e.g.,][]{LyndenBell.Kalnajs.72, Tremaine.Weinberg.84, Carlberg.Sellwood.85, Weinberg.89, Weinberg.91, Weinberg.04, Kaur.Sridhar.18, Banik.vdBosch.21a, Kaur.Stone.21, Chiba.Schonrich.21}, often in the context of phenomena like angular momentum transport, radial migration or dynamical friction.

\subsection{Perturbation in an isothermal slab}
\label{sec:iso_slab}

The infinite slab has a non-uniform (uniform) density profile along the vertical (horizontal) direction. Therefore the unperturbed motion of the stars is only vertically bounded by a potential but is unbounded horizontally. This implies that the unperturbed DF, $f_0$, involves a potential $\Phi_z$ only along $z$. For simplicity, we assume it to be isothermal but with different velocity dispersions in the vertical direction, $\sigma_z$, and the in-plane directions, $\sigma_x = \sigma_y \equiv \sigma$, i.e.,
\begin{align}
f_0(v_x,v_y,E_z)= \frac{\rho_c}{{(2\pi)}^{3/2}\sigma_z\,\sigma^2} \, \exp\left[-\frac{E_z}{\sigma^2_z}\right] \, \exp\left[-\frac{v^2_x+v^2_y}{2\sigma^2}\right],
\label{f_iso}
\end{align}
where 
\begin{align}
E_z = \frac{1}{2} v^2_z +\Phi_z(z)
\label{E_z}
\end{align}
is the energy involving the $z$-motion. The corresponding density and potential profiles in the vertical direction are given by
\begin{align}
\rho_z(z) = \rho_c \, {\sech}^2(z/h_z),\;\;\;\;\;\;\;\;\;\;\;\Phi_z(z) = 2 \sigma^2_z \,\ln\left[\cosh(z/h_z)\right],
\end{align}
where $h_z$ is the vertical scale height \citep[][]{Spitzer.42,Camm.50}. The vertical action, $I_z$, can be obtained from the unperturbed Hamiltonian, $E_z$, as follows
\begin{align}
I_z = \frac{1}{2\pi} \oint v_z \, d z= \frac{2}{\pi} \int_0^{z_{\rm max}} \sqrt{2[E_z - \Phi_z(z)]} \, d z,
\end{align}
where $\Phi_z(z_{\rm max})=E_z$, i.e., $z_{\rm max} = h_z \cosh^{-1}\left(\exp{\left[E_z/2\sigma^2_z\right]}\right)$. The time period of vertical oscillation is given by
\begin{align}
T_z = \oint \frac{d z}{v_z} = 4\int_0^{z_{\rm max}} \frac{d z} {\sqrt{2\left[E_z-\Phi_z(z)\right]}},
\label{T_z}
\end{align}
and the vertical frequency is $\Omega_z = 2\pi/T_z$. Throughout this paper, to compute the perturbative response of the slab, we shall use typical MW parameter values, i.e., $h_z=0.4$ kpc, $\sigma_z=23$ km/s, and $\sigma=1.5\,\sigma_z=35$ km/s \citep[][]{McMillan.11}.

Substituting the above form for $f_0$ (Equation~[\ref{f_iso}]) in Equation~(\ref{f1nk_slabsol}) and using that $\Omega_z = \Omega_z(I_z) = \partial E_z/\partial I_z$ yields the following closed integral form for $f_{1nk}$:
\begin{align}
f_{1nk}(I_z,v_x,v_y,t) &= -i\left(\frac{n\Omega_z}{\sigma^2_z} + \frac{k v_x}{\sigma^2}\right) \, f_0(v_x,v_y,E_z) \, \int_{t_i}^{t}d \tau \, \exp{\left[-i(n\Omega_z+k v_x)(t-\tau)\right]} \, \Phi_{nk}(I_z,\tau).
\label{f1nk_isosol}
\end{align}

\subsection{Perturber potential}
\label{sec:per_pot}

The slab response depends on the spatio-temporal nature of the perturber. In this paper we consider two different functional forms of the perturber potential described below.

\subsubsection{Separable potential}
\label{sec:sep_per_pot}

In order to capture the essential physics of perturbative collisionless dynamics without much computational complexity, we shall consider the following separable form for the perturber potential:
\begin{align}
\Phi_\rmP(z,x,t) = \Phi_\rmN\, \calZ(z) \calX(x) \calT(t),
\label{Phip_sep}
\end{align}
where $\Phi_\rmN$ has the units of potential, and $\calZ$, $\calX$ and $\calT$ are dimensionless functions of $z$, $x$ and $t$ respectively that specify the spatio-temporal profile of $\Phi_\rmP$. Thus, the Fourier transform of $\Phi_\rmP$ can also be written in the following separable form,
\begin{align}
\Phi_{nk}(I_z,t) = \Phi_\rmN\, \calZ_n(I_z) \calX_k\, \calT(t).
\label{Phip_sep_fourier}
\end{align}
Here $\calZ_n(I_z)$ is the $n^{\rm th}$ Fourier coefficient in the discrete Fourier series expansion of $\calZ(z)$ in the vertical angle, $w_z$, given by
\begin{align}
\calZ_n(I_z) = \frac{1}{2\pi} \int_0^{2\pi} d w_z \, \calZ(z) \, \exp{\left[-in w_z\right]},
\label{Phip_sep_fourier_z}
\end{align}
where we have used the implicit expression for $z$ in terms of $w_z$ and $I_z$ given in equation~(\ref{z_wz_Iz}). $\calX_k$ is the Fourier transform of $\calX(x)$, given by
\begin{align}
\calX_k = \frac{1}{2\pi} \int_{-\infty}^{\infty} d x\,  \calX(x) \, \exp{\left[-ikx\right]}.
\label{Phip_sep_fourier_x}
\end{align}
In the following sections, we investigate the slab response to perturbers with various functional forms for $\calX(x)$ and $\calT(t)$, while keeping the form for $\calZ(z)$ arbitrary. We start in Section~\ref{sec:impulsive_kick} with an impulsive ($\calT(t)=\delta(t)$) single-mode ($\calX(x) = \exp[ikx]$) perturbation, followed in Section~\ref{sec:localized} by a perturbation that is temporally impulsive but spatially localized ($\calX(x)=\exp{\left[-x^2/\Delta^2_x\right]}$). In Section~\ref{sec:non-impulsive} we consider the same spatially localized perturbation, but this time temporally extended ($\calT(t)=\exp{\left[-\omega^2_0 t^2\right]}$).

\subsubsection{Satellite galaxy along straight orbit}
\label{sec:sat_per_pot}

As a practical astrophysical application of our perturbative formalism, we also study the response of an isothermal slab to a satellite galaxy or DM subhalo undergoing an impact along a straight orbit with a uniform velocity $\vp$ at an angle $\thetap$ (with respect to the disk normal). We model the impacting satellite as a point perturber, whose potential is given by

\begin{align}
\Phi_\rmP(z,x,t) = -\frac{GM_\rmP}{\sqrt{{\left(z-\vp\cos{\thetap} t\right)}^2+{\left(x-\vp\sin{\thetap} t\right)}^2}}.
\label{Phip_sat}
\end{align}
In this case the spatial and temporal parts are coupled and thus the slab response needs to be evaluated by performing the $\tau$ integral before the $w_z$ and $x$ integrals (to find $\Phi_{nk}$), as shown in Appendix~\ref{App:sat_disk_resp}.

\section{Response to an Impulsive Perturbation}
\label{sec:impulsive_kick}

In order to gain some insight into the perturbative response of the slab, we start by solving equation~(\ref{f1nk_isosol}) for an instantaneous impulse at $t=0$; i.e., $\calT(t) = \delta(t)$. Here the normalization factor $\Phi_\rmN$ has the units of potential times time. With the initial time $t_i<0$, the integral over $\tau$ yields $\exp{\left[-i(n\Omega_z+k v_x)t\right]}$. Further integrating $f_{1nk}$ over $v_x$ and $v_y$ and summing over all $n$ modes, yields the following form for any given $k$ mode of the perturbed DF for a given action $I_z$ and angle $w_z$,
\begin{align}
f_{1k}(I_z,w_z,t)&=\sum_{n=-\infty}^{\infty}\exp{\left[in w_z\right]}\int_{-\infty}^{\infty}d v_y\int_{-\infty}^{\infty}d v_x\, f_{1nk}(I_z,v_x,v_y,t)\nonumber \\
&=A_{\rm norm}\, D_k(t)\, R_k(I_z,w_z,t),
\label{f1_delta}
\end{align}
where 

\begin{align}
A_{\rm norm}=\frac{\rho_c}{\sqrt{2\pi}\sigma_z} \exp{\left[-E_z/\sigma^2_z\right]}
\end{align}
is a normalization factor reflecting the vertical structure of the unperturbed disk,

\begin{align}
D_k(t)=\exp{\left[-\frac{k^2\sigma^2t^2}{2}\right]}
\end{align}
is a damping term that describes the temporally Gaussian decay of the response by lateral mixing, and

\begin{align}
R_k(I_z,w_z,t)=-\Phi_\rmN\calX_k\sum_{n=-\infty}^{\infty}\calZ_n(I_z)\left(k^2t+i\frac{n\Omega_z}{\sigma^2_z}\right)\exp{\left[i n \left(w_z - \Omega_z\,t\right)\right]}
\label{Rk_impulse}
\end{align}
is a (linear) response function that includes vertical phase-mixing.

Equation~(\ref{f1_delta}) is the basic `building block' for computing the response of our infinite isothermal slab to a perturbation mode $k$ in the impulsive limit. Using the canonical transformation from $(w_z,I_z)$ to $(z,v_z)$, i.e., using equations~(\ref{z_wz_Iz}) and~(\ref{E_z}), we can transform $f_{1k}(I_z,w_z,t)$ to $f_{1k}(v_z,z,t)$. Upon multiplying this by $\exp{\left[ikx\right]}$ and integrating over $k$, and then integrating further over $v_z$ at a fixed $z$, one obtains the response density as a function of both time and position:
\begin{align}
\rho_1(z,x,t) &=-\frac{\rho_c \Phi_\rmN}{\sqrt{2\pi}\sigma_z} \sum_{n=-\infty}^{\infty} \int_0^{\Tilde{I}_z} d I_z\, \frac{\Omega_z}{\sqrt{2\left[E_z-\Phi_z(z)\right]}} \exp{\left[-E_z/\sigma^2_z\right]} \exp{\left[i n \left(\Tilde{w}_z - \Omega_z\,t\right)\right]} \,\calZ_n(I_z) \nonumber \\
&\times \int d k\,\exp{\left[ikx\right]}\,  \exp{\left[-\frac{k^2\sigma^2t^2}{2}\right]}\left(k^2t+i\frac{n\Omega_z}{\sigma^2_z}\right)\calX_k,
\end{align}
where $\Tilde{I}_z$ is the solution of $E_z(I_z)=\Phi_z(z)$, and $\Tilde{w}_z$ is the solution for $w_z(z,I_z)$ from equation~(\ref{z_wz_Iz}).

\begin{figure*}[t!]
  \centering
  \includegraphics[width=0.95\textwidth]{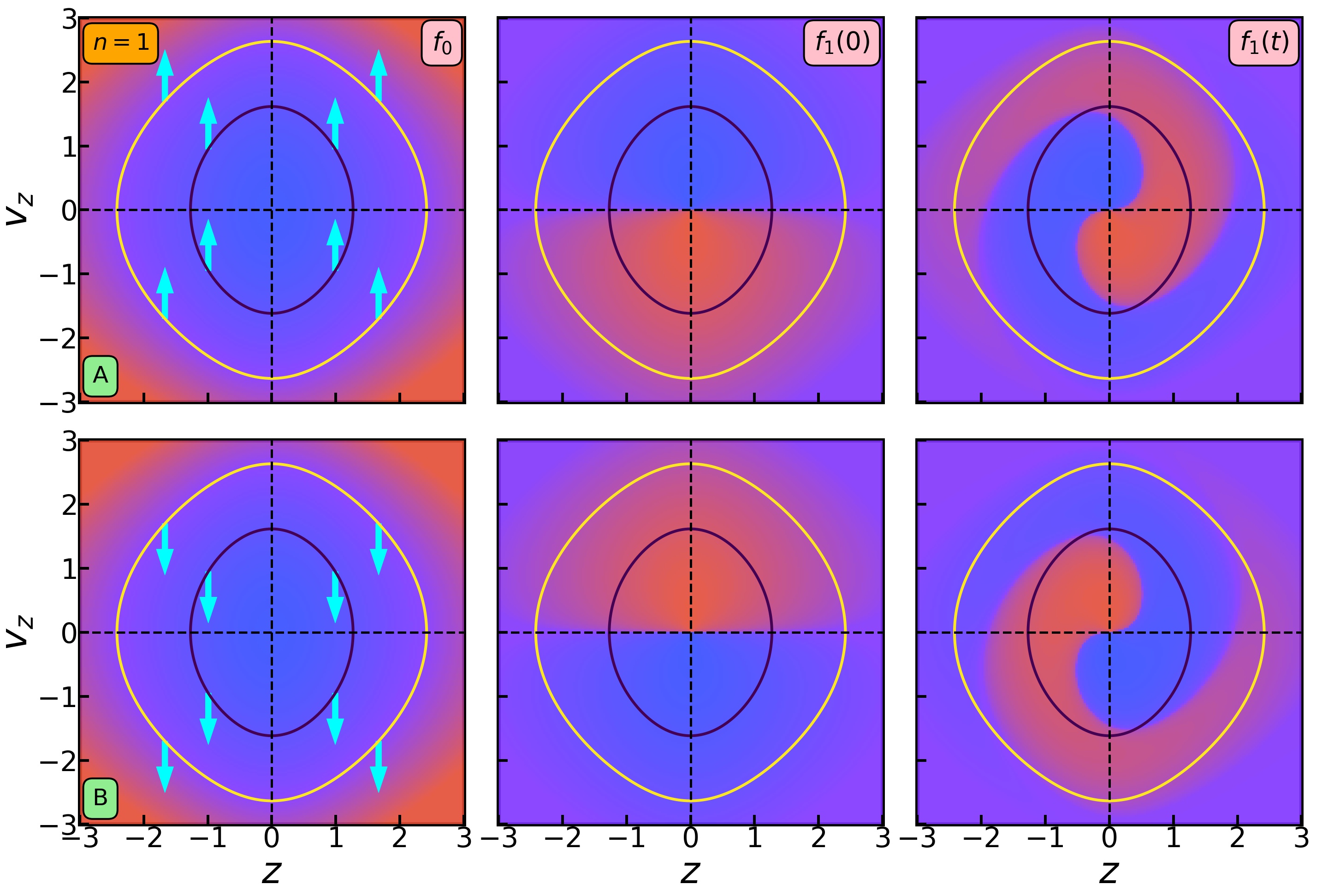}
  \caption{The formation of a one-armed phase spiral due to an impulsive $n=1$ bending-mode perturbation. The color-coding in the left-hand panels shows the unperturbed distribution function $f_0(z,v_z)$ (equation~[\ref{f_iso}]) in the isothermal slab at neighboring locations A (top) and B (bottom), separated by a lateral distance of $\pi/k$, with blue (red) indicating a higher (lower) phase-space density. Locations A and B coincide with extrema in the perturbation mode as depicted in Fig.~\ref{fig:modes}. The black and yellow contours indicate the phase-space trajectories for two random values of $E_z$ (or, equivalently, $I_z$). The cyan arrows indicate the velocity impulses resulting from the instantaneous perturbation at different locations in phase-space. Note that, in the case of the $n=1$ mode considered here, at the extrema A and B all velocity impulses $\Delta v_z$ are positive and negative, respectively (cf. Fig~\ref{fig:modes}). The middle panels indicate the response $f_1$ immediately following the instantaneous response (at $t=0$), with blue (red) indicating a positive (negative) response density. Finally, the right-hand panels show the response after some time $t$, computed using equation~(\ref{f1_delta}). Note how the response at A reveals a one-armed phase spiral that is exactly opposite of that at location B, i.e., they exactly cancel each other. Hence, lateral mixing causes damping of the phase spiral amplitude.}
  \label{fig:impulsive_slab_n1}
\end{figure*}

In order to gain insight into the slab response for a particular $I_z$ and $w_z$, let us start by analyzing equation~(\ref{f1_delta}) for the $n=0$ mode, an in-plane density wave, for which the perturbation causes an in-plane velocity impulse as depicted in Fig.~\ref{fig:modes}. The response is a standing, longitudinal oscillation in density. The response function for this mode is $R_k(I_z,w_z,t) = \Phi_\rmN \calZ_0(I_z)\calX_k\, k^2 t$, indicating that the amplitude of oscillation initially grows linearly with time.  However, this growth is inhibited by the Gaussian damping function $D_k(t) = \exp[-\half k^2 \sigma^2 t^2]$, which describes lateral mixing due to the non-zero velocity dispersion of stars in the $k$-direction. The Gaussian form of this temporal damping term owes its origin to the assumed Gaussian/Maxwellian form of the unperturbed velocity distribution along the plane. Hence, following the perturbation, the $n=0$ mode starts to grow linearly with time, but then rapidly damps away; the response loses its coherence due to mixing in the direction of the wave-vector. In the cold slab limit $(\sigma\to 0)$, without any lateral streaming motion to damp it out, the response will grow linearly in time until it eventually becomes non-linear. This is because in an infinite, laterally homogeneous slab there is no restoring force in the lateral directions, causing the stars to stream uninhibited towards (away from) the minima (maxima) of $\Phi_\rmP$ due to the initial velocity impulse induced. This leads to over- and under-density spikes which cannot be treated using linear theory. Hence, Equation~(\ref{f1_delta}) can only adequately describe the response to an instantaneous $n=0$ mode at early times, or if the damping time $\tau_\rmD \sim (k \sigma)^{-1}$ is shorter than the time-scale of formation of density spikes. The latter is roughly the time needed to cross one quarter of the perturbation's wavelength with the velocity impulse triggered at the zeroes of $\Phi_\rmP$. Therefore, in order for linear theory to be valid, we require that $\sigma > (2/\pi) \, |\Delta v|_{\rm max}$, where $|\Delta v|_{\rm max}=k\, \Phi_\rmN \calZ_0(I_z)\calX_k$. Moreover, upon including self-gravity, it can be found that the $n=0$ mode is linearly stable only for $k>k_J\approx \sqrt{4\pi G \rho_c}/\sigma$ \citep[][]{Binney.Tremaine.08}, or in other words $\lambda<\lambda_J\approx \sigma\sqrt{\pi/G \rho_c}$, where $k_J$ and $\lambda_J=2\pi/k_J$ refer to the Jeans wave-number and Jeans wavelength respectively. In the $\sigma\to 0$ limit, the Jeans wave-length, $\lambda_J\to 0$, and thus the $n=0$ mode becomes globally unstable. Hence, the condition of Jeans stability requires an additional constraint on $\sigma$: $\sigma>\sqrt{4\pi G\rho_c}/k$. The validity of linear perturbation theory thus requires that for each $k$,

\begin{align}
\sigma > \max{\left[\frac{\sqrt{4\pi G \rho_c}}{k},\frac{2k}{\pi}\Phi_\rmN \calZ_0(I_z)\calX_k\right]}.
\end{align}

\begin{figure*}[t!]
  \centering
  \includegraphics[width=0.95\textwidth]{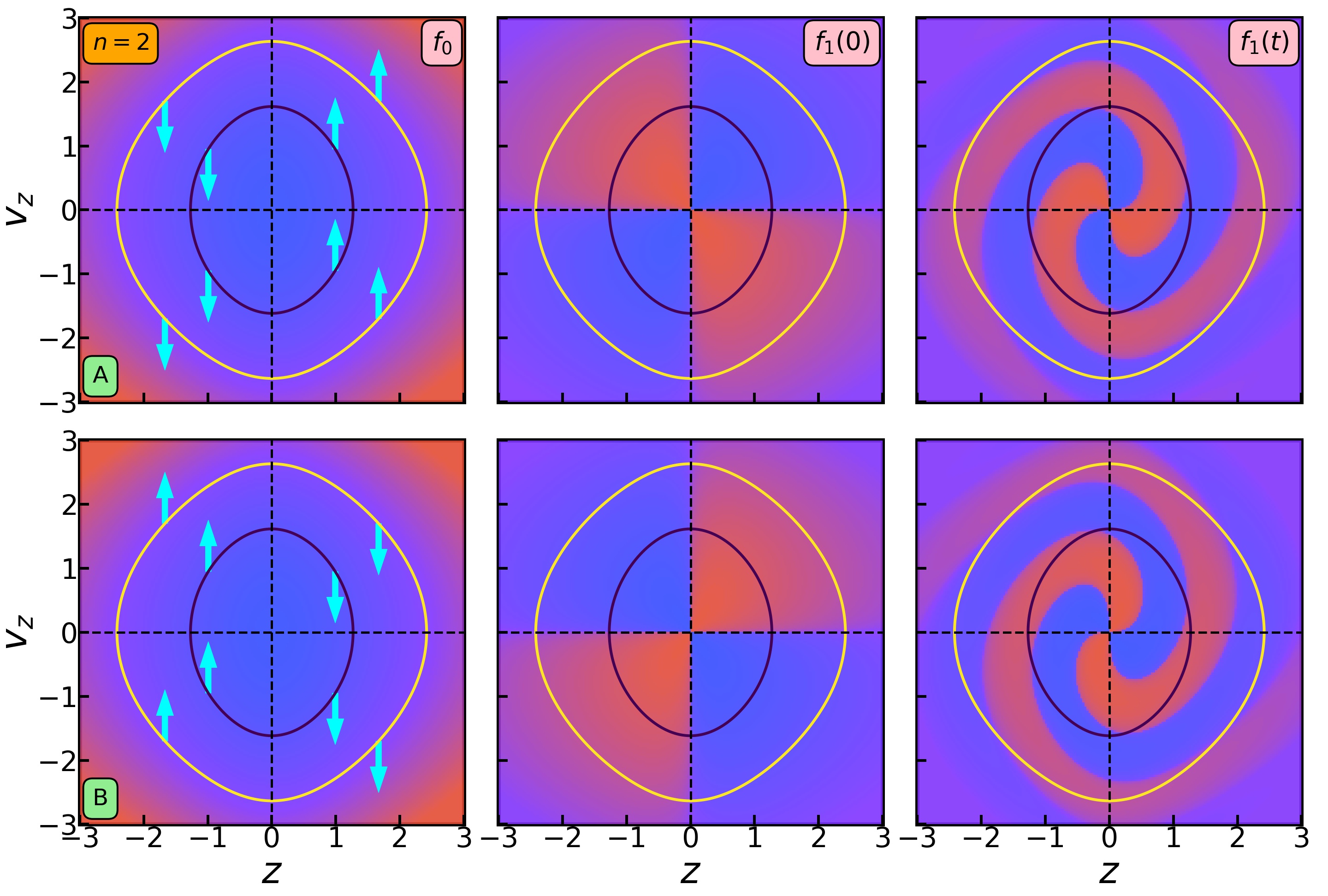}
  \caption{Same as Fig.~\ref{fig:impulsive_slab_n1}, except for a pure $n=2$ breathing mode. Note how in this case the velocity impulses above and below the mid-plane are of opposite sign (cyan arrows in left-hand panels). As a consequence, the response density immediately following the perturbation has a quadrupole signature (middle panels), which ultimately gives rise to two-armed phase spirals (right-hand panels). Note how once again, the phase spirals at A and B are each other's additive inverse.}
  \label{fig:impulsive_slab_n2}
\end{figure*}

For $n=1$, the perturbation is a standing, transverse wave on the slab, formally known as the bending wave. The perturbation induces velocity impulses in the direction perpendicular to the slab, as indicated in  Fig.~\ref{fig:modes}. At the locations marked A and B, separated by a lateral distance of $\pi/k$, these velocity impulses point in the positive and negative $z$-directions, respectively. The top panels of Fig.~\ref{fig:impulsive_slab_n1} illustrate the impact this has at location A. The left-hand panels indicate the velocity impulses (cyan arrows) in the $(z,v_z)$-plane. Prior to the perturbation, due to the vertical restoring force from the slab, each star executes a periodic oscillation in this plane. The black and yellow contours indicate the corresponding phase-space trajectories for two values of $I_z$, while the heat-map indicates phase-space density (bluer colors indicate higher density). The top-middle panel shows that immediately following the impulse, the phase-space density is boosted (reduced) where $v_z>0$ ($v_z<0$), resulting in a dipole pattern in the phase-space distribution of stars. After the impulse, the stars continue to execute periodic motion in the $(z,v_z)$-plane, but starting from their new position (corresponding to a modified action $I_z$). The angular frequency of this periodic motion is $\Omega_z$, which is a function of the (modified) action, and hence, stars with different actions oscillate in the $(z,v_z)$-plane at different frequencies. As a consequence, the perturbed phase-space density shown in the middle panels is wound-up into a {\it phase spiral} of over- and under-densities as depicted in the right-most panels of Fig.~\ref{fig:impulsive_slab_n1}. The bottom panels of Fig.~\ref{fig:impulsive_slab_n1} show what happens following the impulsive perturbation at location B. Since the velocity impulses are now reversed in direction, the phase spiral that emerges is exactly the opposite of that at location A.

The creation of phase spirals is an outcome of phase-mixing in the $z$-direction and is described by the oscillatory factor, $\exp[i\,n(w_z -\Omega_z t)]$, that is part of the response function $R_k(I_z,w_z,t)$. It consists of two terms: a term that scales as $k^2 t$, which describes the lateral streaming motion of stars due to the non-zero velocity impulses in the lateral directions (see Fig.~\ref{fig:modes}), and a term that scales as $n\Omega_z/\sigma^2_z$ which purely describes the vertical oscillations. As in the case of the $n=0$ mode, the lack of a restoring force in the lateral directions\footnote{If accounting for self-gravity of the response density, there will be non-zero forces in the lateral direction, but these will promote growth rather than act as a restoring force. This ultimately leads to exponential growth (according to linear theory) of unstable modes and Landau damping of stable modes, which occurs exponentially, i.e., more slowly than the Gaussian lateral mixing in the absence of self-gravity.} causes the perturbation to grow linearly with time in the absence of lateral streaming (for a cold disk with $\sigma \approx 0$). Meanwhile, the phase spirals continue to wind-up, which implies that the vertical bending loses its coherence. Over time, phase-mixing in the vertical direction will ensure that the disk regains mirror-symmetry with respect to the midplane, but with a scale-height, $h_z$, that would be a periodic function of $x$, with a wavelength equal to $\pi /k$ (i.e., half the wave-length of the original perturbation).
 
However, all this ignores lateral mixing due to the unconstrained motion with non-zero velocity dispersion in the $x$ direction. Stars that received an impulse $\Delta v_z>0$ create phase spirals that are exactly the inverse of those created by neighboring stars for which the impulse was negative. Thus lateral mixing between neighboring points on the slab causes a damping of the phase spiral amplitude at any location, a process that is captured by the damping function $D_k(t)$. The lateral mixing timescale is $\tau_\rmD \sim 1/k\sigma$, indicating, as expected, that small scale perturbations (larger $k$) mix faster, and that mixing is more efficient for larger velocity dispersion in the lateral direction. After a few mixing time-scales, the slab will once again be completely homogeneous (laterally), with a scale-height $h_z$ that is independent of location. In addition, the density of stars in the $(z,v_z)$-plane will once again be perfectly symmetric without any trace of a phase spiral. The slab has completely equilibrated, and the only impact that remains of the impulsive perturbation is that the new scale-height is somewhat larger than it was originally, i.e., the impulsive perturbation has injected energy into the disk, which causes it to puff-up in the vertical direction. Hence, the final outcome is as envisioned in the impulsive-heating scenario discussed in the seminal study of \citet{Toth.Ostriker.92}. This persistent effect in the vertical density profile is however only captured in perturbation theory at second order \citep[e.g.,][]{Carlberg.Sellwood.85}; to first order the perturbation simply phase-mixes away in the impulsive limit considered here.

For $n=2$, the perturbation triggers a breathing mode, as depicted in Fig.~\ref{fig:modes}, i.e., at a given location A on the slab, the velocity impulses for this mode are positive (negative) for positive (negative) $z$. As evident from Fig.~\ref{fig:impulsive_slab_n2}, this leads to a quadrupole pattern for the initial perturbed phase-space distribution of stars, which becomes a two-armed phase spiral over time, as opposed to the one-armed phase spiral resulting from the $n=1$ mode. This reveals an important lesson: the structure of a phase spiral depends, among others, on which perturbation mode(s) are triggered. The phase spirals in regions A and B are each other's additive inverse. Hence, once again lateral mixing will cause damping of the phase spiral's amplitude, as described by the damping function $D_k(t)$. \cite{Hunt.etal.21} have shown using N-body simulations that two-armed phase spirals can indeed arise from breathing mode oscillations and that both bending and breathing modes can be excited at different locations on the MW disk by satellite-induced perturbations such as the passage of Sagittarius (see section~\ref{sec::MW_satellites} for detailed discussion).

To summarize, we see that, in case of our infinite slab, equilibration after an impulsive perturbation is driven by a combination of phase-mixing in the vertical direction and free-streaming damping in the horizontal direction. While the former gives rise to phase spirals in the $(\sqrt{I_z} \cos w_z, \sqrt{I_z} \sin w_z)$ or equivalently the $(z,v_z)$ plane, the latter causes them to damp away by lateral mixing. Due to vertical phase-mixing the phase spiral will continue to wrap itself up into a more and more tightly wound pattern, until its structure can no longer be discerned observationally due to finite-$N$ noise and measurement errors in the actions and angles of individual stars (this is an example of coarse-grain mixing). Hence, even without lateral mixing phase spirals are only detectable for a finite duration.

\section{Response to a localized perturbation}
\label{sec:localized}

In the previous section we investigated the slab response to an external disturbance with a single wavenumber $k$. Realistic perturbations are however localized in space and thus consist of many wavenumbers. In this section we shall look into what happens when the slab is hit by an impulsive perturbation that is spatially localized. 

For simplicity, we assume that the external perturber behaves as a Gaussian packet with half-width $\Delta_x$ along the $x$ direction, i.e., $\Phi_\rmP$ is given by equation~(\ref{Phip_sep}) with
\begin{align}
\calX(x) = \exp{\left[-x^2/2\Delta_x^2\right]}.
\end{align}
The $\calZ(z)$ term in equation~(\ref{Phip_sep}) denotes the vertical structure of the perturber potential, which is part of what dictates the relative strength of bending and breathing mode oscillations. We shall see in the next section, though, that the relative strength of the modes is mostly dictated by the form of $\calT(t)$. For simplicity, we only consider localization along the $x$ and $z$-directions; along the $y$-direction the perturbation is assumed to extend out to infinity. We emphasize, though, that this assumption does not impact the essential physics of phase-mixing and lateral mixing discussed below.

The Fourier transform of the perturber potential, $\Phi_{nk}$, is given by equation~(\ref{Phip_sep_fourier}), with
\begin{align}
\calX_k = \frac{\Delta_x}{\sqrt{2\pi}}\, \exp{\left[-k^2\Delta_x^2/2\right]}.
\label{Phink_gaussian}
\end{align}
Upon substituting the above expression for $\calX_k$ in equation~(\ref{f1_delta}) we obtain the response for a single $k$ mode, $f_{1k}$. After multiplying this by $\exp{\left[ikx\right]}$, integrating over all $k$ and summing over all $n$ modes, we obtain the following final form for the slab response density in the case of a (laterally) Gaussian perturber,
\begin{align}
f_1(I_z,w_z,x,t) &= \sum_{n=-\infty}^{\infty} \exp{\left[i n w_z\right]} \int_{-\infty}^{\infty} d k\, \exp{\left[ikx\right]}\, f_{1k}(I_z,w_z,t) \nonumber \\
&=A_{\rm norm}\, D(x,t)\, R(I_z,w_z,x,t),
\label{f1_local_impulsive}
\end{align}
where 
\begin{align}
A_{\rm norm}=\frac{\rho_c}{\sqrt{2\pi}\sigma_z} \exp{\left[-E_z/\sigma^2_z\right]}
\end{align}
is the same normalization factor as in equation~(\ref{f1_delta}),
\begin{align}
D(x,t)=\frac{\Delta_x}{\sqrt{\Delta_x^2+\sigma^2 t^2}} \exp{\left[-\frac{x^2}{2\left(\Delta_x^2+\sigma^2 t^2\right)}\right]}
\end{align}
is a factor that captures the decay of the response by lateral mixing, and
\begin{align}
R(I_z,w_z,x,t)=-\Phi_\rmN \sum_{n=-\infty}^{\infty}\calZ_n(I_z)\left[\frac{t}{\Delta_x^2+\sigma^2 t^2}\left(1-\frac{x^2}{\Delta_x^2+\sigma^2 t^2}\right)+i\frac{n\Omega_z}{\sigma^2_z}\right]\exp{\left[i n \left(w_z - \Omega_z\,t\right)\right]}
\label{R_local_impulsive}
\end{align}
with $\calZ_n(I_z)$ given by equation~(\ref{Phip_sep_fourier_z}), corresponds to the remaining part of the response that includes vertical phase-mixing.

The above expression (equation~[\ref{f1_local_impulsive}]) for the slab response to a localized disturbance has several important features. Firstly, the profile of the slab response is nearly Gaussian in $x$ since we assumed a Gaussian form (along $x$) for the perturber potential. Secondly, the $D(x,t)$ factor describes the decay of the response amplitude and widening of the response profile due to mixing by lateral streaming. The mixing in this case occurs as a power law in time rather than like a Gaussian as for a single $k$ mode (see equation~[\ref{f1_delta}]), since the power spectrum of the Gaussian perturber is dominated by small $k$ which mix very slowly, at a timescale, $\tau_\rmD\sim 1/k\sigma$. Thirdly, the $R$ factor captures two important effects: (i) a transient response reflecting an initial linear growth due to the perturber-induced velocity impulse, followed by a subsequent decay by lateral mixing, and (ii) vertical oscillations of stars (for $n\neq 0$) at different frequencies resulting in phase-mixing over time and the formation of phase spirals as described in detail in Section~\ref{sec:impulsive_kick}. The $n=0$ modes, i.e., perturbations confined to the slab, damp out faster than the non-zero $n$ modes that manifest the vertical oscillations of stars. Since the perturber was introduced impulsively by means of a Dirac delta function in time, the higher order oscillations are stronger for the same value of $\calZ_n(I_z)$ as the corresponding changes in the vertical actions have larger amplitude. Typically, for $n\geq 2$, $\calZ_n(I_z)$ gets smaller with larger $n$; hence the $n=2$ breathing mode turns out to be the dominant mode of oscillation for impulsive disturbances. The response characteristics however change as we move to non-impulsive or more temporally extended perturbers in the next section.

It takes time for the local response to propagate along the slab by lateral streaming. Initially the perturber's gravity draws in stars towards the center of impact, $x=0$. Thus, immediately following the impulse, the region near the center of impact has a larger concentration of stars, which laterally stream outwards due to non-zero velocity dispersion. This leads to a damping of the response amplitude at small $x$ and growth at large $x$, or equivalently damping and widening of the response profile, which occurs at the rate,
\begin{align}
\calD_x(t) = \frac{d}{d t} \sqrt{\Delta^2_x + \sigma^2 t^2} = \frac{\sigma^2 t}{\sqrt{\Delta^2_x + \sigma^2 t^2}}.
\end{align}
This rate of outward streaming of slab material is initially equal to
\begin{align}
\lim_{t\to 0} \calD_x(t) = \frac{\sigma^2 t}{\Delta_x},
\end{align}
but at later times asymptotes to a constant value,
\begin{align}
\lim_{t\to \infty} \calD_x(t) = \sigma.
\end{align}

To summarize, the response to a spatially localized perturbation can be understood in the context of that to a single mode plane wave perturbation discussed in the previous section, as follows. In both cases, the response involves vertical oscillations that phase-mix away, thus giving rise to phase spirals. However, whereas the plane wave response maintains its sinusoidal profile in the lateral direction with an overall Gaussian decay of the amplitude due to lateral mixing, the response profile in the case of localized perturbation changes its shape and undergoes both decay and widening. This is because in the latter case the response is a linear superposition of responses to many plane wave perturbations with different $k$, each decaying in amplitude over a time-scale, $\tau_\rmD\sim 1/k\sigma$, due to lateral mixing by free-streaming. Since the spatially Gaussian profile considered here has a Gaussian power spectrum and thus more power on large scales (small $k$) that mix more slowly, the combined response from all $k$ modes undergoes much slower lateral mixing (as a power law) than that from a single $k$ mode. The typical timescale of coarse-grained survival (against free-streaming damping) of the phase spiral in this case turns out to be $\sim (f_{\rm max}/f_{\rm res})\,\Delta_x/\sigma$. Here $f_{\rm max}$ is the maximum amplitude of the phase spiral, which is attained at $t=0$, and $f_{\rm res}$ is the resolution limit. The power law nature of free-streaming damping implies that the response to spatially and temporally localized perturbations (e.g., encounters with satellite galaxies) can be sustained in the disk for a long time.

\section{Response to a non-impulsive perturbation}
\label{sec:non-impulsive}

Thus far we have only considered impulsive perturbations of our slab, with $\calT(t)=\delta(t)$. However, a realistic disturbance would not only have a spatial structure, the effects of which we studied in the previous section, but also be extended in time. In this section we investigate the effect of non-impulsive or temporally extended disturbances on the slab oscillations. In particular, we broaden the Dirac delta pulse from the previous section into a Gaussian in time, i.e., $\Phi_\rmP$ is given by equation~(\ref{Phip_sep}) with $\calT(t) = \frac{1}{\sqrt{\pi}}\,\exp{\left[-\omega^2_0 t^2\right]}$, where $\omega_0$ is the pulse frequency. We define the pulse-width or pulse-time as $\tau_\rmP=\sqrt{2}/\omega_0$. We also assume that the pulse is localized and follows a Gaussian profile in $x$ as in the previous section, i.e., $\calX(x)=\exp{\left[-x^2/2\Delta^2_x\right]}$. As before, $\calZ(z)$ in equation~(\ref{Phip_sep}) denotes some generic vertical profile. The (spatial) Fourier transform of this potential, $\Phi_{nk}$, is provided in equation~(\ref{Phip_sep_fourier}) with $\calX_k$ given by equation~(\ref{Phink_gaussian}) and $\calZ_n$ given by equation~(\ref{Phip_sep_fourier_z}). We can substitute this in equation~(\ref{f1nk_slabsol}) and perform the integration over $\tau$ and $v_x$ to obtain the following expression for the response for a single $k$ mode,
\begin{align}
&f_{1k}(I_z,w_z,t)=A_{\rm norm}\, D_k(t)\,R_k(I_z,w_z,t),
\label{f1k_gaussian}
\end{align}
where 
\begin{align}
A_{\rm norm}=\frac{\rho_c}{\sqrt{2\pi}\sigma_z} \exp{\left[-E_z/\sigma^2_z\right]}
\end{align}
is the same normalization factor as in equation~(\ref{f1_delta}),

\begin{align}
D_k(t) = \frac{\calQ^3}{2\omega_0}\,  \exp{\left[-\calQ^2\frac{k^2\sigma^2t^2}{2}\right]}
\end{align}
is a factor that describes the damping of the response due to lateral mixing, and

\begin{align}
&R_{k}(I_z,w_z,t)=-\Phi_\rmN\calX_k \sum_{n=-\infty}^{\infty} \calZ_n(I_z) \left\{\,
S_{nk} \, \Upsilon_{nk}(t) \, \left(k^2 t + i \frac{n\Omega_z}{\sigma^2_z}\right) \exp{\left[i\, n(w_z-\calQ\,\Omega_z t)\right]} - \calG_{nk}(w_z,t)\right\},
\label{Rk_gaussian}
\end{align}
with $\calZ_n(I_z)$ given by equation~(\ref{Phip_sep_fourier_z}), includes the vertical phase-mixing of the response. Here $\calQ$ is a factor that depends on the pulse frequency, $\omega_0$, and the wavenumber $k$, and is given by
\begin{align}
\calQ=\calQ(\omega_0,k\sigma)=\frac{\omega_0}{\sqrt{\omega^2_0+\frac{k^2\sigma^2}{2}}}.
\end{align}
The mode-strength,
\begin{align}
S_{nk} = \exp{\left[-\frac{1}{\omega^2_0+\frac{k^2\sigma^2}{2}}\frac{n^2\Omega^2_z}{4}\right]}
\label{mode_strength_gaussian}
\end{align}
is a function that indicates the strength of each $n$ mode,
\begin{align}
\Upsilon_{nk}(t) &=  1+\erf\left\{\calQ \left(\omega_0 t-i\frac{n\Omega_z}{2\omega_0}\right)\right\}
\label{growth_gaussian}
\end{align}
describes the temporal build-up of the response and the decay of transient oscillations, and
\begin{align}
\calG_{nk}(w_z,t) = \frac{k^2}{\sqrt{\pi}\,\omega_0\calQ} \exp{\left[-\calQ^2\omega^2_0 t^2\right]} \exp{\left[in w_z\right]}
\label{transient_gaussian}
\end{align}
is another rapidly decaying transient feature. In the $\omega_0 \to \infty$ limit, both $\Upsilon_{n}(t)$ and the mode strength $S_{nk}$ become unity, and $\calG_{nk}(w_z,t) \to 0$, such that we recover the response to impulsive perturbations given in equation~(\ref{f1_delta}) as required.

It is interesting to contrast this response to an extended pulse to that in the impulsive limit. First of all, the damping factor, $D_k(t)$, which still owes its origin to lateral mixing due to non-zero velocity dispersion, now depends not only on $k$ and $\sigma$ but also on the pulse frequency $\omega_0$. The damping time is given by
\begin{align}
\tau_{\rmD} = \frac{1}{k\sigma} \sqrt{1+\frac{k^2\sigma^2}{2\omega^2_0}},
\end{align}
which scales as $\sim 1/k\sigma$ in the impulsive/short pulse ($\omega^2_0 \gg k^2\sigma^2/2$) limit indicating that the response mixes away laterally with small scale perturbations mixing faster. In the adiabatic/long pulse ($\omega^2_0 \ll k^2\sigma^2/2$) limit, though, $\tau_\rmD \to 1/\sqrt{2}\omega_0$, i.e., the damping of the response follows the temporal behaviour of the perturbing pulse itself, independent of $k$.

The mode-strength reveals several important trends: it exponentially damps away with $n^2$, implying that the lower order modes are much stronger for perturbations that are slower \citep[see also][]{Widrow.etal.14} and/or have larger wavelength (smaller $k$). Therefore the $n=1$ bending modes dominate over the $n=2$ breathing modes for a sufficiently slow pulse. Note, though, that if the pulse is too slow ($\omega_0 \to 0$) the mode strength is super-exponentially suppressed, especially at large scales (small $k$), or if the slab has a small lateral velocity dispersion, $\sigma$, compared to that along the vertical direction, $\sigma_z$ (recall that $\Omega_z\sim \sigma_z/h_z$). This is a classic signature of adiabatic shielding of the slab due to the averaging out of the net response to zero by many oscillations of stars within the (very long) perturbation timescale \citep[cf.][]{Weinberg.94a,Weinberg.94b,Gnedin.Ostriker.99}.

\begin{figure*}
  \centering
  \includegraphics[width=0.95\textwidth]{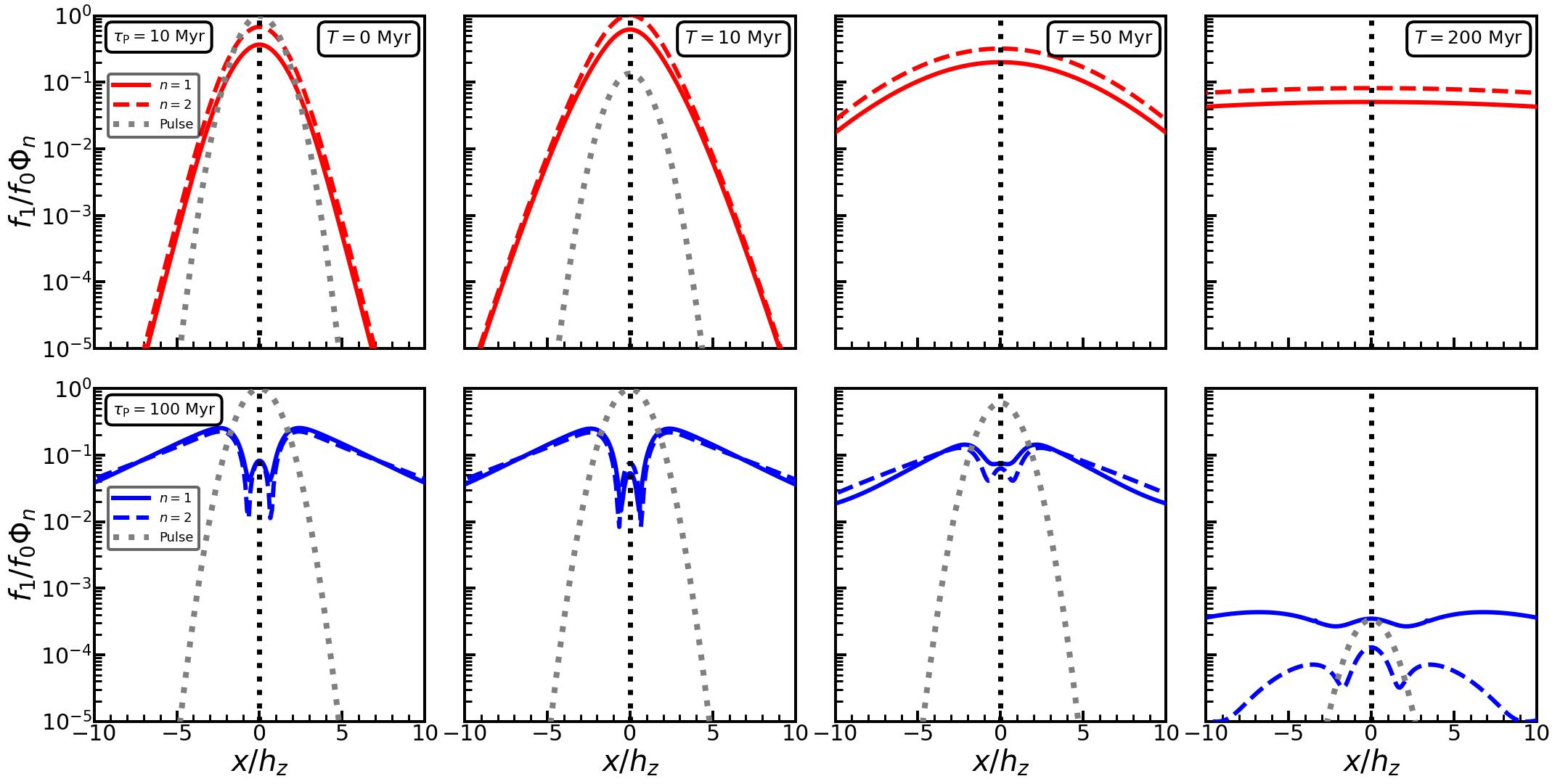}
  \caption{Amplitude of the slab response to a Gaussian (in both $x$ and $t$) packet of half-width $\Delta_x=h_z$ as a function of $x$ for different times since the maximum pulse-strength. The two rows indicate two different pulse times, as indicated. We adopt our fiducial MW parameters (see Section~\ref{sec:iso_slab}) and take $I_z=0.5\, h_z\sigma_z$. Solid (dashed) lines show $n=1$ ($n=2$) bending (breathing) modes, while the grey-dotted lines show the perturbing pulse, $\calT(t)\calX(x)$. The response density initially grows and then damps away due to lateral mixing. In the short pulse limit, the response density is Gaussian in $x$, which damps out and widens like a power law in time. The response in the longer pulse behaves like a sinusoid at small $x$ (see Appendix~\ref{App:ad_lim_resp}) and its intensity shows a transient growth followed by exponential damping before it falls off as a power law. The bending (breathing) mode eventually dominates in the slow (fast) pulse limit.}
  \label{fig:gaussian_slab_x}
\end{figure*}

Finally, if the perturbation is not impulsive the frequency with which the slab stars oscillate in the vertical direction is modified with respect to their natural frequency according to
\begin{align}
\Omega_z \to \frac{\omega^2_0}{\omega^2_0+\frac{k^2\sigma^2}{2}} \Omega_z,
\end{align}
which goes to $\Omega_z$ in the impulsive limit, as expected. For slower pulses however, the vertical motion of the stars couples to the lateral motion \citep[see also][]{Binney.Schonrich.18}, resulting in a reduced oscillation frequency, especially for smaller wavelengths (larger $k$). In the extremely slow/adiabatic limit, $\Omega_z \to 0$, signalling a lack of vertical phase-mixing. This is easy to understand; a forced oscillator remains in phase with the perturber if the frequency of the latter is much lower than the natural frequency. In fact, in the adiabatic limit, the response only consists of resonant stars, for which $n\Omega_z+k v_x=0$ (see Appendix~\ref{App:ad_lim_resp}), and thus no phase spiral emerges.

\begin{figure*}
  \centering
  \includegraphics[width=1\textwidth]{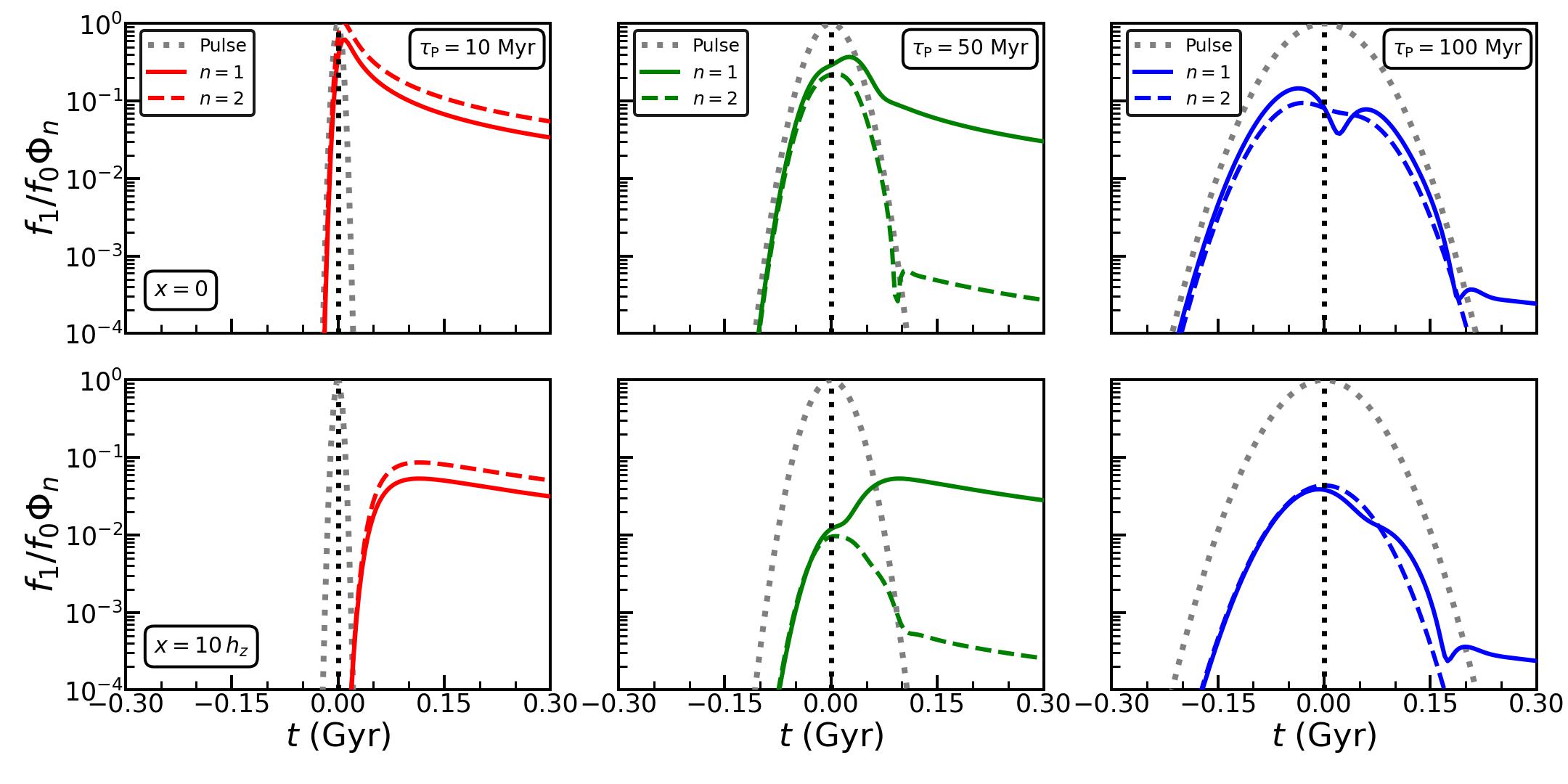}
  \caption{Amplitude of the slab response to a Gaussian perturbation (in both $x$ and $t$) at two locations in the slab: at the location of impact, $x = 0$, shown in the top panels, and at a distance $x = 10 h_z$ away, shown in the bottom panels. As in Fig.~\ref{fig:gaussian_slab_x}, the spatial Gaussian wave-packet, $\calX(x)$, has a half-width of $\Delta_x = h_z$.  Different columns correspond to different values of the Gaussian pulse-widths, $\tau_\rmP$, as indicated. The grey-dotted line in each panel shows the perturbing pulse $\calT(t)$ at $x=0$, while solid and dashed lines show responses for the $n=1$ (bending) and $n=2$ (breathing) modes. The response to shorter pulses shows a transient growth followed by a power law fall-off with time. Response to longer pulses initially grows and then damps away as a Gaussian before finally transitioning to a power law fall-off. For longer pulses, the bending modes dominate in the long run, while for shorter pulses, the breathing modes are stronger.}
  \label{fig:gaussian_slab_t}
\end{figure*}

The above response corresponds to a temporally Gaussian pulse for a fixed wavenumber $k$. To get the full response to a localized perturber, we substitute the expression for $\calX_k$ given in equation~(\ref{Phink_gaussian}), in the $k$-response of equation~(\ref{f1k_gaussian}), multiply it by $\exp{\left[ikx\right]}$ and integrate over all $k$. The resultant response is an oscillating function of $w_z$ and has a profile along $x$ which varies with time. For the short pulse/impulsive case, we recover the expression given in equation~(\ref{f1_local_impulsive}). In Fig.~\ref{fig:gaussian_slab_x} we plot the amplitude (relative to the unperturbed DF) of this oscillating response (normalized by the $z$ Fourier component of the perturber potential, $\calZ_n$) as a function of $x$. The columns correspond to four different times since the time of maximum pulse strength, and the rows correspond to two different pulse-times, as indicated. The solid and dashed lines indicate the bending ($n=1$) and breathing ($n=2$) modes, respectively. The short pulse response shown in the upper panels has a Gaussian profile centered on the point of impact at $x=0$ with the initial width very similar to that of the $\Phi_\rmP$ profile (see equations~[\ref{f1_local_impulsive}]-[\ref{R_local_impulsive}]). Over time, this response profile gets weaker and wider like a power law, as the unconstrained lateral motion of the stars causes an outward streaming, and thus decay, of the response. The long pulse response in the lower panels has a different, more extended profile than in the short pulse case; it exhibits some ripples along $x$ besides having an overall smooth behaviour (see Appendix~\ref{App:ad_lim_resp} for the response derived in the adiabatic limit). As time goes on, the response decays away and widens out due to lateral mixing. Unlike the short pulse case, here the response initially decays like $\sim \exp{\left[-\omega^2_0 t^2\right]}$ over a timescale of the pulse-time, $\tau_\rmP=\sqrt{2}/\omega_0$, before attaining a power law decay at large time.

The temporal behaviour of the response becomes even clearer in Fig.~\ref{fig:gaussian_slab_t}, where we plot the amplitude of the response as a function of time at two different positions on the slab (different rows), and for three different pulse-times (different columns). As before, the solid and dashed lines indicate the $n=1$ and $n=2$ modes, respectively. Initially the slab response grows nearly hand in hand with the perturbing pulse. This is captured by the $\Upsilon_{nk}(t)$ term (equation~[\ref{growth_gaussian}]) in the expression for $R_k(I_z,w_z,t)$, which scales as $\exp{\left[-\calQ^2 \omega^2_0 t^2\right]}$ at small $t$, but asymptotes to a constant value of $2$ at late times. As the perturber strength falls off, the response decays as a Gaussian for each $k$, as described by the damping factor, $D_k(t) \propto \exp[-{\calQ}^2k^2\sigma^2 t^2/2]$. The combined response from all $k$ however decays at a different rate. For the shortest pulse, for which the response asymptotes to that given by equation~(\ref{f1_local_impulsive}), the damping factor, $D(x,t) \propto 1/t$ at late times. In the intermediate and long pulse cases, the response initially tends to follow the same $\sim \exp{\left[-\omega^2_0 t^2\right]}$ decay as the perturbing pulse, before finally transitioning to a power law fall-off, which typically occurs as $\sim 1/t$, just as in the short pulse case. Importantly, this transition sets in later for longer lasting pulses, such that the late-time response for slower perturbations is drastically suppressed with respect to faster perturbations. From the bottom panels, it is evident that the region ($x=10 h_z$) farther away from the center of impact responds later, with a time-lag of $\Delta t= 10\, h_z/\sigma$ (timescale of lateral streaming), which is $\sim 115$ Myr for the typical MW parameter values adopted here. The breathing mode is the dominant mode in the short pulse case ($\tau_\rmP = 10 \Myr$) while in both the intermediate ($\tau_\rmP = 50 \Myr$) and long ($\tau_\rmP = 100 \Myr$) pulse scenarios the bending mode eventually dominates. Note, though, that if the pulse becomes too long, the long-term response is adiabatically suppressed. Hence, there is only a narrow window in pulse-widths for which bending modes dominate and cause a detectable response. In the next section we examine whether any of the MW satellites have encounters with the disk over timescales that fall in this regime.

The response formalism for localized, non-impulsive perturbations developed so far can be used to model the response to transient bars and spiral arms. Encounters with such features can cause transient vertical perturbations in the potential over timescales comparable to the vertical oscillation periods of stars, thereby creating phase-spirals. We discuss this in detail in Paper~II for realistic disk galaxies.

\section{Encounters with satellite galaxies}
\label{sec:sat_encounter}

In all cases considered above we have made the simplifying assumption that the perturber potential is separable, i.e., can be written in the form of equation~(\ref{Phip_sep}). However a realistic perturber is seldom of such simple form. For example, the potential due to an impacting satellite galaxy or DM subhalo (approximated as a point perturber) cannot be written in separable form, thereby making the analysis significantly more challenging. In this section, as an astrophysical application of the perturbative formalism developed in this paper, we compute the response of the infinite slab to a satellite encounter. We relegate the far more involved computation of the response of a realistic disk to an impacting satellite to Paper~II.

As shown in Appendix~\ref{App:sat_disk_resp}, the $n\neq 0$ response to a satellite impacting the slab with a uniform velocity $\vp$ along a straight orbit at an angle $\thetap$, at a distance $x$ from the point of impact, can be approximated as
\begin{align}
f_1(I_z,w_z,x,t) = \frac{\rho_c}{\sqrt{2\pi}\sigma_z} \exp{\left[-E_z/\sigma^2_z\right]}\times i\frac{2G\Mp}{\vp} \sum_{n=-\infty}^{\infty} \frac{n\Omega_z}{\sigma^2_z}\, \Psi_n(x,I_z)\, \exp{\left[i\,\frac{n\Omega_z \sin{\thetap}}{\vp}x\right]} \exp{\left[i n\left(w_z-\Omega_z t\right)\right]},
\label{f1_sat}
\end{align}
where
\begin{align}
\Psi_n(x,I_z)&= \frac{1}{2\pi} \int_0^{2\pi} d w_z\, \exp{\left[-i n \left(w_z - \frac{\Omega_z \cos{\thetap} z}{\vp}\right)\right]} K_0\left[\,\left|\frac{n\Omega_z \left(x\cos{\thetap}-z\sin{\thetap}\right)}{\vp}\right|\,\right],
\label{Psi_n}
\end{align}
with $K_0$ the zero-th order modified Bessel function of the second kind. This expression for the response is only valid far away from the point of impact ($x\gtrsim \sigma t$), such that the response can be approximated as a plane wave along $x$, and at late times, after the perturber has moved far enough away from the disk, i.e., for $t \gg (x\sin{\thetap}+z\cos{\thetap})/\vp$).

There are several salient features of this response that deserve special attention. The strength of the response is dictated by the $K_0$ function whose argument depends on $\Omega_z \cos{\thetap}\, x/\vp$ (for small $I_z$), which is basically the ratio of the encounter timescale,
\begin{align}
\tau_{\rm enc} = \frac{x\cos{\thetap}}{\vp},
\label{tau_enc}
\end{align}
and the vertical dynamical time of the stars,
\begin{align}
\tau_z = \frac{1}{\Omega_z} \sim \frac{h_z}{\sigma_z}.
\label{tau_z}
\end{align}
From the asymptotic limits of $K_0$ it follows that the response scales as a power law ($\sim \vp^{-1}$) in the impulsive ($\tau_{\rm enc}\ll \tau_z$) limit and as $\sim\exp{\left[-\left|n\Omega_z\cos{\thetap}\right|x/\vp\right]}$ in the adiabatic ($\tau_{\rm enc}\gg \tau_z$) limit. The response peaks roughly at the maximum of the $K_0$ function, which occurs when the encounter timescale is comparable to the vertical dynamical time of the stars, i.e., when $\tau_{\rm enc}\approx 0.6\,\tau_z$, or in other words the `resonance' condition,
\begin{align}
\frac{x\cos{\thetap}}{\vp} \approx \frac{0.6}{\Omega_z},
\end{align}
is satisfied. For encounters faster than this, the response is suppressed like a power law, while for slower encounters it is exponentially suppressed. The $\vp^{-1}$ scaling of the response in the impulsive limit is a well known feature of impulsive perturbations \citep[e.g.,][]{Spitzer.58, Aguilar.White.85, Weinberg.94a, Weinberg.94b, Gnedin.etal.99, Banik.vdBosch.21b}, and the exponential suppression is a telltale signature of adiabatic shielding\footnote{While the adiabatic response in one degree-of-freedom cases, e.g., the vertical phase spiral in the isothermal slab, is exponentially suppressed, that in multiple degree-of-freedom systems such as inhomogeneous disks is usually not because of resonances \citep[][]{Weinberg.94a,Weinberg.94b}.}, similar to the adiabatic suppression of the mode-strength factor in the response to slow Gaussian pulses discussed in section~\ref{sec:non-impulsive}.

While the response is heavily damped for very slow encounters, something interesting happens in the mildly slow regime, $\tau_{\rm enc}=x\cos{\thetap}/\vp \gtrsim \tau_z$. In this regime, the ratio of the $n=2$ breathing to the $n=1$ bending mode response scales as
\begin{align}
f_{21} \equiv {f_{1,n=2} \over f_{1,n=1}} \sim \sqrt{2}\,\exp{\left[-\frac{\Omega_z\cos{\thetap}\,x}{\vp}\right]}.
\label{f21}
\end{align}
Thus the bending mode response exponentially dominates over that of the breathing mode for slower (smaller $\vp$), more distant (large $x$), and more perpendicular ($\thetap\approx 0$) encounters. The bending mode is also more pronounced for stars with larger $\Omega_z$ or equivalently smaller $I_z$. On the other hand, for encounters with $\tau_{\rm enc}=x\cos{\thetap}/\vp < \tau_z$, the breathing modes dominate.

Finally, the slab response to the impacting satellite, given in equation~(\ref{f1_sat}), consists of oscillating functions of time, lateral distance $x$, and the vertical oscillation amplitude, $\sqrt{2I_z/\nu}$ (see equations~[\ref{Psi_n_app_epi}] and [\ref{Phin_sat_app}]). This implies that the satellite not only induces temporal oscillations, which give rise to phase-mixing and thus phase spirals due to different oscillation frequencies of the stars (see section~\ref{sec:impulsive_kick}), but also spatial corrugations. These vertical and lateral waveforms have wavenumbers given by
\begin{align}
k_z = \frac{n\Omega_z\cos{\thetap}}{\vp},\;\;\;\;\;\;\;{\rm and}\;\;\;\;\;\;\;\;
k_x = \frac{n\Omega_z\sin{\thetap}}{\vp},
\end{align}
respectively. Thus, perpendicular impacts induce only vertical corrugations while planar ones excite waves only laterally. An inclined encounter, on the other hand, spawns corrugations along both directions. Both wavelengths get longer with decreasing mode order, increasing impact velocity, and decreasing vertical frequencies, i.e., increasing actions.

\begin{figure*}
  \centering
  \begin{subfigure}{0.35\textwidth}
    \centering
    \includegraphics[width=1\textwidth]{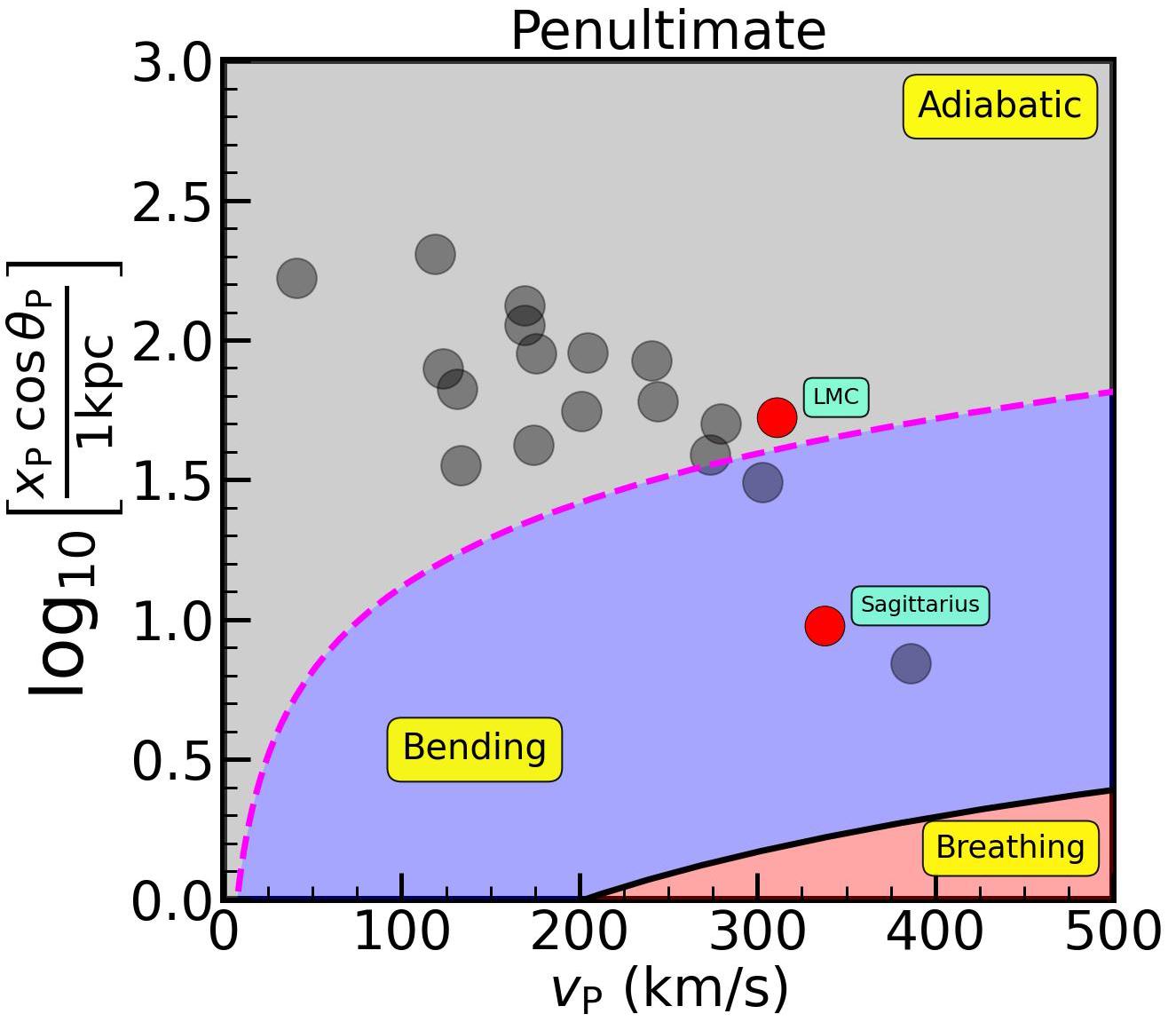}
    \label{disk_resp_n1_0.5pi}
  \end{subfigure}
  \begin{subfigure}{0.63\textwidth}
    \centering
    \includegraphics[width=1\textwidth]{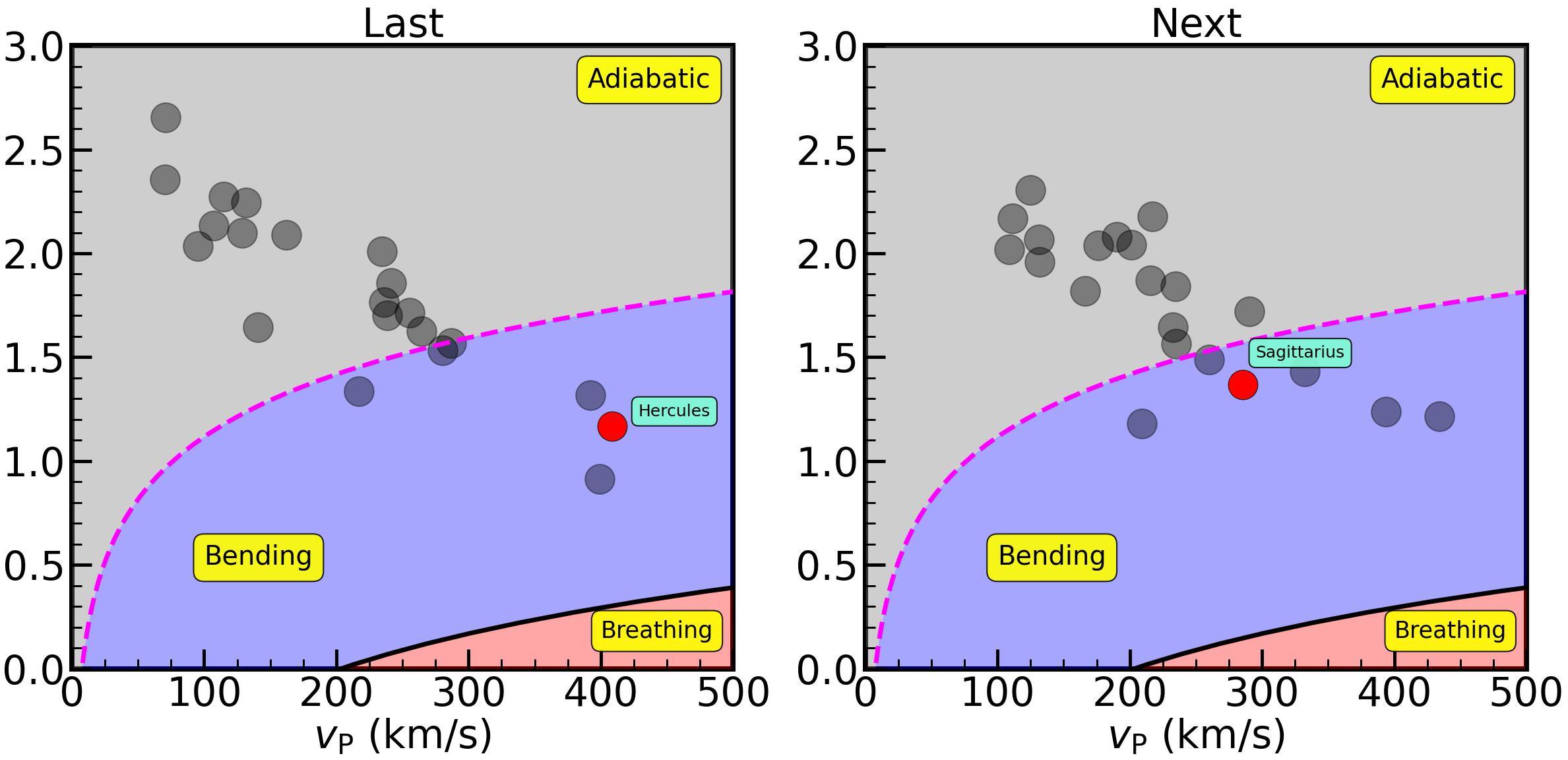}
    \label{disk_resp_n1_0.5pi}
  \end{subfigure}
  \caption{Regions in the space of impact parameter, $\xp\cos{\thetap}$, and velocity, $\vp$, of a satellite galaxy, corresponding to bending (blue) and breathing (red) mode responses in the Solar neighborhood. Response is adiabatically suppressed in the grey region. The circles in the left, middle and right panels indicate the values of $\xp\cos{\thetap}$ and $\vp$ for several MW satellites during their penultimate, last and next disk crossings respectively. The satellites that induce a relative bending mode response, $f_{1,n=1}/f_0\gtrsim 10^{-4}$, for $I_z=h_z\sigma_z$ in the Solar neighborhood, are indicated by red circles, while the others are denoted in grey. All the MW satellites lie outside the breathing region and thus preferentially excite bending modes in the vicinity of the Sun.}
  \label{fig:satellite_constraints1}
\end{figure*}

\subsection{Impact of satellite galaxies on the Milky Way disk}
\label{sec::MW_satellites}

The MW halo harbors many satellite galaxies. Some of these are quite massive, with DM halo mass comparable to the disk mass, and either underwent or are about to undergo an encounter with the MW disk within a few hundred Myr from the present day. Hence we expect at least some of them to perturb the disk significantly. Here we use existing data on MW satellites to obtain a rough estimate of the disk response to their encounters with the MW stellar disk.

Our formalism provides physical insight into the trends and scalings of the disk response as a function of impact parameters and velocities of the MW satellites. We emphasize upfront, though, that the precise numerical estimates of the responses are to be taken with a grain of salt. These estimates only serve as a crude, order-of-magnitude attempt to compare the relative disk responses to different satellite galaxies. As discussed in more detail in section~\ref{sec::caveats}, these estimates are subject to a number of oversimplifications and caveats. First of all, the MW disk is modelled as an isothermal slab, and we only consider the {\it direct} impact of the satellites. We ignore indirect effects due to the self-gravity of the response. Our approach also ignores the presence of a dark matter halo, which can impact the disk response in several ways (see section~\ref{sec::caveats}). Because of all these shortcomings, we caution against using the following response estimates for comparison with actual data and/or detailed numerical simulations.

\begin{figure*}
  \centering
  \includegraphics[width=1\textwidth]{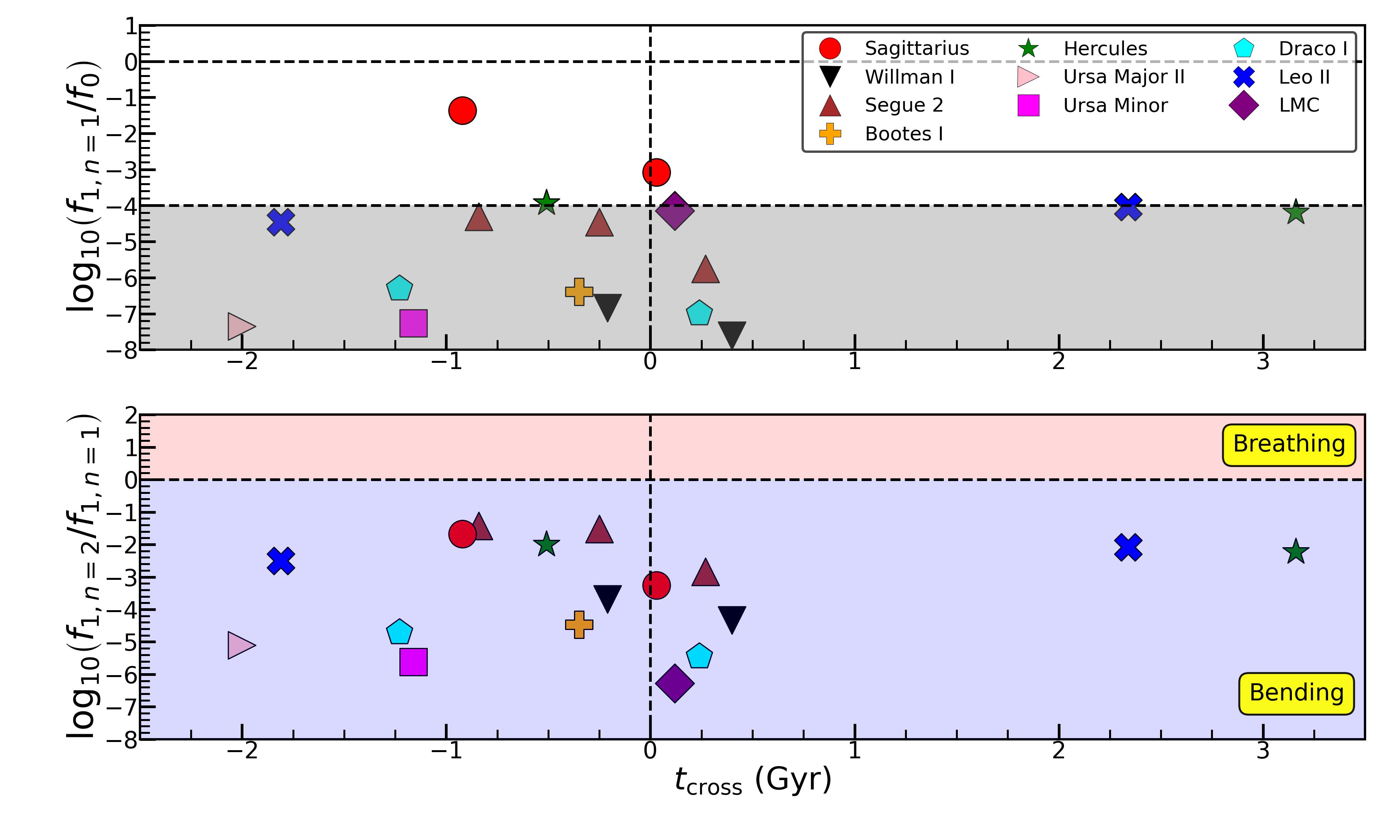}
  \caption{Bending mode strength, $f_{1,n=1}/f_0$ (upper panel), and the corresponding breathing vs bending ratio, $f_{1,n=2}/f_{1,n=1}$ (lower panel), in the Solar neighborhood for the MW satellites, as a function of the disk crossing time, $t_{\rm cross}$, in Gyr, where $t_{\rm cross}=0$ marks today. The previous two and the next impacts are shown. Here we consider $I_z=h_z \sigma_z$, with fiducial MW parameters. In the upper panel, the region with bending mode response, $f_{1,n=1}/f_0<10^{-4}$, has been grey-scaled, indicating that the response from the satellites in this region is far too adiabatic and weak. Note that the response is dominated by that due to Sgr, followed by Hercules, Leo II, Segue 2 and the Large Magellanic Cloud (LMC). Also note that the previous two and next impacts of all the satellites shown here excite bending modes in the Solar neighborhood.}
  \label{fig:satellite_constraints2}
\end{figure*}

We consider the MW satellites with parallax and proper motion measurements from Gaia DR2 \citep[][]{Gaia_collab_sat.18b} and the corresponding galactocentric coordinates and velocities computed and documented by \cite{Riley.etal.19} \citep[table A.2, see also][]{Li.etal.20} and \cite{Vasiliev.Belokurov.20}. Of these, we only consider the satellites with known dynamical mass estimates \citep[][]{Simon.Geha.07,Bekki.Stanimirovic.09,Lokas.09,Erkal.etal.19}. Adopting the initial conditions for galactocentric positions ($R,z,\phi$) and velocities ($v_R,v_z,v_\phi$) as the median values quoted by \cite{Riley.etal.19} and \cite{Vasiliev.Belokurov.20}, we simulate the orbits of the galaxies in the combined gravitational potential of the MW halo, disk and bulge, which are respectively modelled by a spherical NFW \citep[][]{Navarro.etal.97} profile (virial mass $M_h=9.78\times 10^{11}\Msun$, scale radius $r_h=16$ kpc, and concentration $c=15.3$), a Miyamoto-Nagai \citep[][]{Miyamoto.Nagai.75} profile (mass $M_d=9.5\times 10^{10} \Msun$, scale radius $a=4$ kpc, and scale-height $b=0.3$ kpc), and a spherical \cite{Hernquist.90} profile (mass $M_b=6.5\times 10^9\Msun$ and scale radius $r_b=0.6$ kpc)\footnote{Our MW potential is similar to {\tt GALPY MWPOTENTIAL2014} \citep[][]{Bovy.15} except for the power-law bulge which has been replaced by an equivalent Hernquist bulge.}. The total mass of our fiducial MW model is thus $1.08\times 10^{12}\Msun$. We evolve the positions and velocities of the satellites both forwards and backwards in time from the present day, using a second order leap-frog integrator. For simplicity, we ignore the effect of dynamical friction\footnote{Dynamical friction might play an important role in the orbital evolution of massive satellites like the Large Magellanic Cloud (LMC) and Sgr, pushing their orbital radius farther out in the past.}. From each individual orbit, we note the time, $t_{\rm cross}$, when the satellite crosses the disk (i.e., crosses $z=0$), and record the corresponding distance, $\xp$, from the Sun, which we integrate backwards/forwards in time using a purely circular orbit up to $t_{\rm cross}$. We also record the velocity, $\vp=\sqrt{v^2_R+v^2_z+v^2_\phi}$, and the angle of impact with respect to the disk normal, $\thetap=\cos^{-1}{(v_z/\vp)}$. Finally, we compute the disk response to the satellite encounter using equation~(\ref{f1_sat}). Results are summarized in Table~\ref{tab:MW_sat_resp} and Figs.~\ref{fig:satellite_constraints1} and~\ref{fig:satellite_constraints2}.

%

%
%

In Fig.~\ref{fig:satellite_constraints1}, we plot the impact parameter, $\xp\cos{\thetap}$ (with respect to the Sun), as a function of the encounter velocity, $\vp$, of the satellites, for the penultimate (left-hand panel), last (middle panel), and next (right-hand panel) disk crossings. The red (grey) symbols denote the satellites that induce a strong (weak) amplitude of the bending mode response, $f_{1,n=1}/f_0$, for $I_z=h_z\sigma_z = 9.2 \kpc\kms$. As shown in Appendix~\ref{App:detect_crit}, we consider $f_{1,n=1}/f_0=\delta=10^{-4}$ as a rough estimate for the minimum detectable relative response, i.e., the boundary between strong and weak responses to satellite passage. The solid black line indicates the boundary between bending and breathing modes, i.e., where the breathing-to-bending ratio, $f_{21}$ (equation~[\ref{f21}]), is equal to unity. Hence, the blue and red shaded regions indicate where the response is dominated by bending and breathing modes, respectively. The magenta, dashed line roughly denotes the boundary between a strong bending response (blue shaded region) and a response that is adiabatically suppressed (grey shaded region). The latter is defined by the condition $\exp{\left[-\Omega_z \xp \cos{\thetap}/\vp\right]} < \delta = 10^{-4}$.

In Fig.~\ref{fig:satellite_constraints2}, we plot the amplitude of the bending mode response, $f_{1,n=1}/f_0$ (upper panel), and the breathing-to-bending ratio, $f_{21}=f_{1,n=2}/f_{1,n=1}$ (lower panel), in the Solar neighborhood, as a function of the time $t_{\rm cross}$ (in Gyr) when the satellite crosses the plane of the disk, assuming the fiducial MW parameters. Negative and positive $t_{\rm cross}$ correspond to disk crossings in the past and future, respectively, and we once again consider stars with $I_z = h_z \sigma_z = 9.2 \kpc\kms$.

Both Fig.~\ref{fig:satellite_constraints1} and the lower panel of Fig.~\ref{fig:satellite_constraints2} make it clear that {\it all} the disk crossings considered here preferentially excite bending rather than breathing modes in the Solar neighborhood. As shown in Section~\ref{sec:impulsive_kick} these trigger one-armed phase spirals in the Solar neighborhood, in qualitative agreement with the MW snail observed in the Gaia data. However, as is evident from the upper panel of Fig.~\ref{fig:satellite_constraints2}, most satellites only trigger a minuscule response in the disk, with  $f_{1,n=1}/f_0 < \delta = 10^{-4}$, either because the satellite has too low mass, or because the encounter, from the perspective of the Sun, is too slow such that the local response is adiabatically suppressed.  The strongest response by far is triggered by encounters with Sgr, for which the bending mode response, $f_{1,n=1}/f_0$, is at least $1-2$ orders of magnitude larger than that for any other satellite. Based on our orbit-integration, it had its penultimate disk crossing, which closely coincides with its last pericentric passage, about $900\Myr$ ago, triggering a strong response of $f_{1,n=1}/f_0 \sim 0.04$ in the Solar neighborhood. The last disk crossing, which nearly corresponds to the last apocentric passage, occurred about $300\Myr$ ago, triggering a very weak (adiabatically suppressed) response. Sgr is currently near its pericenter and will undergo the next disk crossing in about $30\Myr$, which we estimate to only trigger a moderately strong response with $f_{1,n=1}/f_0\sim 0.001$. We caution, though, that in addition to the caveats listed above and in Section~\ref{sec::caveats} these estimates ignore dynamical friction and are sensitive to the MW potential and the current phase-space coordinates of the satellites. We have checked that a heavier MW model with a total mass of $1.5\times 10^{12}\Msun$ does not change the relative amplitudes of the satellite responses significantly, but brings most of the disk crossing times closer to the present day since the satellites are more bound in a heavier MW. For example, the previous pericentric and apocentric passages of Sgr occur at $\sim 600$ and $200\Myr$ ago in the heavier case. The only satellite apart from Sgr that triggers a response $f_{1,n=1}/f_0 > \delta = 10^{-4}$ is Hercules, whose disk crossing $\sim 500\Myr$ ago caused a bending-mode response, $f_{1,n=1}/f_0 = 1.2\times 10^{-4}$. Segue 2 induces a response that is marginally below the detection threshold. Disk crossings of LMC and Leo II trigger responses that are comparable in strength to that of Hercules, but the crossing times are too far in the past or future for them to be considered as candidates for triggering the Gaia snail. All in all, it is clear then that Sgr is by far the most likely candidate among the MW satellite galaxies considered here to have triggered the one-armed phase spiral in the Solar neighborhood discovered in Gaia DR2 by \citet{Antoja.etal.18}. 

\begin{table*}
\centering
\hspace{-3cm}
\tabcolsep=0.2 cm
\begin{tabular}{c|c|cc|cc|cc}
 \hline
MW satellite & Mass & $f_{1,n=1}/f_0$ & $t_{\rm cross}$ & $f_{1,n=1}/f_0$ & $t_{\rm cross}$ & $f_{1,n=1}/f_0$ & $t_{\rm cross}$ \\
name & $(\Msun)$ & & $(\Gyr)$ & & $(\Gyr)$ & & $(\Gyr)$ \\
 & & Penultimate & Penultimate & Last & Last & Next & Next \\
 (1) & (2) & (3) & (4) & (5) & (6) & (7) & (8) \\
 \hline
 \highlight{Sagittarius} & $10^9$ & \highlight{$4.3\times 10^{-2}$} & \highlight{$-0.92$} & $1.4\times 10^{-10}$ & $-0.3$ & $8.3\times 10^{-4}$ & $0.03$ \\
 Hercules & $7.1\times 10^6$ & -- & $-3.57$ & $1.2\times 10^{-4}$ & $-0.51$ & $6.4\times 10^{-5}$ & $3.16$ \\
 Leo II & $8.2\times 10^6$ & -- & $-3.61$ & $3.5\times 10^{-5}$ & $-1.81$ & $9.3\times 10^{-5}$ & $2.34$ \\
 Segue 2 & $5.5\times 10^5$ & $5\times 10^{-5}$ & $-0.84$ & $3.4\times 10^{-5}$ & $-0.25$ & $1.8\times 10^{-6}$ & $0.27$ \\
 LMC & $1.4\times 10^{11}$ & $1.4\times 10^{-4}$ & $-6.97$ & -- & $-2.37$ & $7.2\times 10^{-5}$ & $0.12$ \\
 SMC & $6.5\times 10^9$ & $3.6\times 10^{-8}$ & $-3.22$ & -- & $-1.39$ & $1.2\times 10^{-9}$ & $0.22$ \\
 Draco I & $2.2\times 10^7$ & -- & $-2.43$ & $5\times 10^{-7}$ & $-1.23$ & $1\times 10^{-7}$ & $0.24$ \\
 Bootes I & $10^7$ & -- & $-1.65$ & $4.1\times 10^{-7}$ & $-0.35$ & -- & $0.87$ \\
 Willman I & $4\times 10^5$ & -- & $-0.63$ & $1.4\times 10^{-7}$ & $-0.21$ & $2.5\times 10^{-8}$ & $0.4$ \\
 Ursa Minor & $2\times 10^7$ & -- & $-2.26$ & $5.5\times 10^{-8}$ & $-1.16$ & $8.6\times 10^{-9}$ & $0.29$ \\
 Ursa Major II & $4.9\times 10^6$ & $4.5\times 10^{-8}$ & $-2$ & $6.2\times 10^{-10}$ & $-0.1$ & -- & $0.9$ \\
 Coma Berenices I & $1.2\times 10^6$ & $7\times 10^{-10}$ & $-2.47$ & -- & $-0.25$ & -- & $0.69$ \\
 Sculptor & $3.1\times 10^7$ & -- & $-2.7$ & $2\times 10^{-10}$ & $-0.46$ & -- & $1.47$\\
 \hline
\end{tabular}
\caption{MW disk response to satellites for stars with $I_z=h_z\sigma_z$ in the Solar neighborhood. Column (1) indicates the name of the MW satellite and Column (2) indicates its dynamical mass estimate from literature \citep[][]{Simon.Geha.07,Bekki.Stanimirovic.09,Lokas.09,Erkal.etal.19,Vasiliev.Belokurov.20}. We assume $10^9\Msun$ for the Sagittarius mass; note that there is a discrepancy between its measured mass of $\sim 4\times 10^8\Msun$ \citep[][]{Vasiliev.Belokurov.20} and the required mass of $10^9-10^{10}\Msun$ for observable phase spiral signatures in N-body simulations \citep[see for example][]{Bennett.etal.21}. Columns (3) and (4) respectively indicate the bending mode response assuming fiducial MW parameters and the crossing time for the penultimate disk-crossing. Columns (5) and (6) show the same for the last disk-crossing, while columns (7) and (8) indicate it for the next one. Only the satellites that trigger a bending mode response, $f_{1,n=1}/f_0\geq 10^{-10}$, in at least one of the three cases are shown. The responses smaller than $10^{-10}$ are considered far too adiabatic and negligible and are marked by dashes. The case most relevant for the Gaia phase spiral is highlighted in red.}
\label{tab:MW_sat_resp}
\end{table*}

We emphasize that the results shown in Figs.~\ref{fig:satellite_constraints1} and~\ref{fig:satellite_constraints2} correspond to stars with a vertical action $I_z=h_z\sigma_z=9.2\, \kpc\kms$. As mentioned above, the strength of the response depends on the ratio of the encounter time scale, $\tau_{\rm enc}$ (equation~[\ref{tau_enc}]) and the vertical oscillation period of stars in the Solar neighborhood, $\tau_z$ (equation~[\ref{tau_z}]). The latter is longer for stars with larger vertical action, and from the perspective of such stars the encounter is more impulsive, resulting in a stronger response. Since the response does not scale linearly with $\tau_{\rm enc}/\tau_z$, the relative response strength of different satellites depends somewhat on the vertical action. We have verified that for $I_z/(h_z\sigma_z) < 3$, which is roughly the range covered by the Gaia phase spiral, the direct response from the encounter with Sgr remains larger than that of any other satellite considered here by at least $1-2$ orders of magnitude. However, for stars with larger actions (larger vertical excursions), the LMC can dominate the response. In particular, for stars with $I_z/(h_z\sigma_z) \gtrsim 6.5$ ($z_{\rm max}\gtrsim 4\, h_z$), which make up the thick disk, the LMC is expected to trigger a stronger response than Sgr during its upcoming disk crossing.

To summarize, our analysis suggests that the MW satellites during their most recent and forthcoming disk crossings preferentially excite bending modes in the Solar neighborhood. This is because satellite encounters are fairly distant from the Sun and thus the encounter time exceeds the vertical oscillation time of the stars. However, as previously discussed in section~\ref{sec:sat_encounter} and as evident from the N-body simulation of MW-Sgr encounter by \cite{Hunt.etal.21} (especially the earlier disk passages of Sgr), a satellite passage can trigger breathing modes closer to the point of impact, where the encounter is more impulsive. Since almost all the MW satellites undergo their disk-crossings at $R\gg 8\kpc$, future observations of the outskirts of the disk might reveal breathing instead of bending mode oscillations if they are excited by any of the satellites considered here.

\subsection{Caveats}
\label{sec::caveats}

The above calculation of the response of the MW disk to perturbations is subject to a number of oversimplifications and caveats discussed below.

The MW disk is modelled as an isothermal slab, which lacks the axisymmetric density profile and velocity structure that characterize a realistic disk. In particular, whereas the lateral motion in our slab is uninhibited, the in-plane motion in a realistic disk consists of an azimuthal rotation combined with a radial epicyclic motion. Among others, this will have important implications for the global disk response and the rate at which phase spirals damp out due to lateral mixing. In Paper~II (Banik et al., in preparation) we apply our perturbative formalism to a realistic self-gravitating disk galaxy with a pseudo-isothermal distribution function \citep[][]{Binney.10}, and consider both external perturbations (encounters with satellites) and internal perturbations (bars and spiral arms).
    
All responses calculated in this paper only account for the direct response to a perturbing potential. In general, though, the response also has an indirect component that arises from the fact that neighboring regions in the disk interact with each other gravitationally. This self-gravity of the response, which we have ignored, triggers long-lived normal mode oscillations of the slab that are not accounted for in our treatment. Several simulation-based studies have argued that including self-gravity is important for a realistic treatment of phase spirals \citep[e.g.,][]{Darling.Widrow.19a, Bennett.Bovy.21}. Using the Kalnajs matrix method \citep[][]{Kalnajs.77, Binney.Tremaine.08}, we have made some initial attempts to include the self-gravity of the response in our perturbative analysis, along the lines of \citet{Weinberg.91}. Our preliminary analysis shows that the self-gravitating response is a linear superposition of two terms: (i) a continuum of modes given in equation~(\ref{f1nk_slabsol}), dressed by self-gravity, that undergo phase-mixing and give rise to the phase spiral, and (ii) a discrete set of modes called point modes or normal modes \citep[c.f.][]{Mathur.90,Weinberg.91} that follow a dispersion relation. The continuum response can be amplified by self-gravity when the continuum frequencies, $n\Omega_z+k v_x$, are close to the point mode frequencies, $\nu$. Depending on the value of $k$, the normal modes can be either stable or unstable. \cite{Araki.85} find that in an isothermal slab the bending normal mode undergoes fire hose instability below a certain critical wavelength if $\sigma_z/\sigma \lesssim 0.3$ while the breathing normal mode becomes unstable above the Jeans scale. In the regime of stability, the normal modes are undamped oscillation modes in absence of lateral streaming \citep[][]{Mathur.90} but are Landau damped otherwise \citep[][]{Weinberg.91}. For an isothermal slab with typical MW-like parameter values, the point modes are strongly damped since their damping timescale (inverse of the imaginary part of $\nu$) is of order their oscillation period (inverse of the real part of $\nu$), which turns out to be of order the vertical dynamical time, $h_z/\sigma_z$. Moreover, the normal mode oscillations are coherent oscillations of the entire system, independent of the vertical actions of the stars, and are decoupled from the phase spiral in linear theory since the full response is a linear superposition of the two. Based on the above arguments, we conclude that self-gravity has little impact on the evolution of phase spirals in the isothermal slab, at least in the linear regime. We emphasize that \citet{Darling.Widrow.19a}, who found their phase spirals to be significantly affected by the inclusion of self-gravity, assumed a perturber-induced velocity impulse with magnitude comparable to the local velocity dispersion in the Solar neighborhood; hence their results are likely to have been impacted by non-linear effects. Moreover, the self-gravitating response of an inhomogeneous disk embedded in a dark matter halo, as in the simulations of \citet{Darling.Widrow.19a}, can be substantially different from that of the isothermal slab. We intend to include a formal treatment of self-gravity along the lines of \citet{Weinberg.91} in future work.
    
The disk of our MW is believed to be embedded in an extensive dark matter halo, something we have not taken into account. The presence of such a halo has several effects. First of all, the satellite not only perturbs the disk, but also the halo. In particular, it induces both a local wake and a global modal response\footnote{The torque from the local as well as global halo response is responsible for dynamical friction acting on the satellite.} \citep[e.g.,][]{Weinberg.89, Tamfal.etal.21}. The former typically trails the satellite galaxy, and boosts its effective mass by about a factor of two \cite[][]{Binney.Tremaine.08}, which might boost the (direct) disk response by about the same factor. The global halo response is typically dominated by a strong $l=1$ dipolar mode followed by an $l=2$ quadrupolar mode \citep[][]{Tamfal.etal.21}, which might have a significant impact on the disk. The presence of a halo also modifies the total potential. At large disk radii and vertical heights, the halo dominates the potential and will therefore significantly modify the actions and frequencies of the stars, and consequently the shape of the phase spirals. Finally, since the disk experiences the gravitational force of the halo, a (sufficiently massive) satellite galaxy can excite normal mode oscillations of the disk in the halo \citep[see for example][]{Hunt.etal.21}. We intend to incorporate some of these effects of the MW halo in Paper~II.

\section{Conclusion}
\label{sec:concl}

In this paper we have used linear perturbation theory to compute the response of an infinite, isothermal slab to various kinds of external perturbations with diverse spatio-temporal characteristics. Although a poor description of a realistic disk galaxy, the infinite, isothermal slab model captures the essential physics of perturbative response and collisionless equilibration via phase-mixing in the disk, and thus serves as a simple yet insightful case for investigation.

We use a hybrid (action-angle variables in the vertical direction and position-momentum variables in the lateral direction) linear perturbative formalism to perturb and linearize the collisionless Boltzmann equation and compute the response in the distribution function of the disk to a gravitational perturbation. We have considered external perturbations of increasing complexity, ranging from an instantaneous (laterally) plane-wave perturbation (Section~\ref{sec:impulsive_kick}), an instantaneous localized perturbation, represented as a wave-packet (Section~\ref{sec:localized}), a non-impulsive, temporally extended, localized perturbation (Section~\ref{sec:non-impulsive}), and ultimately an encounter with a satellite galaxy moving along a straight-line orbit (Section~\ref{sec:sat_encounter}). This multi-tiered approach is ideal for developing the necessary insight into the complicated response that is expected from a realistic disk galaxy exposed to a realistic perturbation. We summarize our conclusions below.

\begin{itemize}
    \item The two primary Fourier modes of slab oscillation are the $n=1$ bending mode and the $n=2$ breathing mode, which correspond to anti-symmetric and symmetric oscillations about the mid-plane, respectively. For a sufficiently impulsive perturbation, the dominant mode is the breathing mode, which initially causes a quadrupolar distortion in the $(z,v_z)$-phase space, that evolves into a two-armed phase spiral as the stars with different vertical actions oscillate with different vertical frequencies. If the perturbation is temporally more extended (less impulsive), the dominant mode is the bending mode. This causes a dipolar distortion in $(z,v_z)$-phase space that evolves into a one-armed phase spiral \citep[see also][]{Hunt.etal.21, Widrow.etal.14}. Due to vertical phase-mixing, the phase spiral wraps up tighter and tighter until it becomes indistinguishable from an equilibrium distribution in the coarse-grained sense.
    
    \item Besides vertical phase-mixing the survivability of the phase spiral is also dictated by the lateral streaming motion of stars. The initial lateral velocity impulse towards the minima of $\Phi_\rmP$ tends to linearly boost the contrast of the phase spiral. This is however quickly taken over by lateral streaming (with velocity dispersion $\sigma$), which causes mixing between the over- and under-densities, and damps out the phase spiral amplitude. For an impulsive, laterally sinusoidal perturbation, the disk response is also sinusoidal and damps out like a Gaussian (due to the Maxwellian/Gaussian distribution of the unconstrained lateral velocities) over a timescale of $\tau_\rmD \sim 1/k\sigma$, i.e., small scale perturbations damp out faster, as expected.
    
    \item Lateral mixing operates differently for a spatially localized perturbation which can be expressed as a superposition of many plane waves. The response to each of them damps out like a Gaussian (if the perturber is impulsive). Since the power spectrum of a spatially localized perturber with a lateral Gaussian profile is dominated by its largest scales (small $k$) that mix and damp out slower, the net response from all $k$ damps away as $\sim t^{-1}$ (the response profile spreads out as $\sim t$), much slower than the Gaussian damping in case of a sinusoidal perturber.
    
    \item The disk response to a non-impulsive perturbation is substantially different from that to an impulsive one. If the temporal strength of the perturber follows a Gaussian pulse with pulse frequency, $\omega_0$ (e.g., a transient bar or spiral arm), the response grows and decays following the temporal profile of the pulse before eventually attaining a $\sim 1/t$ power law fall-off. The response peaks when the pulse frequency, $\omega_0$, is comparable to the vertical oscillation frequency, $\Omega_z$. The response to more impulsive perturbations ($\omega_0 \gg \Omega_z$) is suppressed as $\sim 1/\omega_0$, whereas much slower ($\omega_0 \ll \Omega_z$) perturbations trigger a super-exponentially ($\sim \exp{\left[-n^2\Omega^2_z/4\omega^2_0\right]}$ at small $k$) suppressed response. In this adiabatic limit, the stars tend to remain in phase with the perturber, oscillating at frequencies much smaller than $\Omega_z$, which inhibits the formation of a phase spiral.
    
    \item The timescale of perturbation dictates the excitability of different modes, with slower (faster) pulses triggering stronger bending (breathing) modes. An encounter with a satellite galaxy that hits the disk with a uniform velocity $\vp$ and an angle $\thetap$ with respect to the normal at a distance $\xp$ away from an observer in the disk, perturbs the potential at an observer's location with a characteristic time scale $\tau_{\rm enc} \sim \xp\cos{\thetap}/\vp$. If $\tau_{\rm enc}$ is long (short) compared to the typical vertical oscillation time, $\tau_z \sim h_z / \sigma_z$, at the observer's location, the dominant perturbation mode experienced is a bending (breathing) mode. Thus, bending modes are preferentially excited not only by low velocity encounters, but also by more distant and more perpendicular ones. Since the velocities of all MW satellites are much larger than $\sigma_z$, the decisive factor for bending {\it vs.} breathing modes is the distance from the point of impact. This is in qualitative agreement with the results from $N$-body simulations of the MW-Sgr encounter performed by \cite{Hunt.etal.21}, which show more pronounced bending (breathing) modes further from (closer to) the location where Sgr impacts the disk. Moreover, for a given encounter, stars with larger actions undergo stronger breathing mode oscillations since they oscillate slower.
    
    \item Besides phase spirals satellite encounters also induce spatial corrugations in the disk response, with vertical and lateral wave-numbers given by $k_z=n\Omega_z\cos{\thetap}/\vp$ and $k_x=n\Omega_z\sin{\thetap}/\vp$, respectively.
\end{itemize}

As an astrophysical application of our formalism, we have investigated the direct response of the MW disk (approximated as an isothermal slab) to several of the satellite galaxies in the halo for which dynamical mass estimates and galactocentric phase-space coordinates from Gaia parallax and proper motion measurements are available. We integrate the orbits of these satellites in the MW potential and note the impact velocity, $\vp$, angle of impact, $\thetap$, with respect to the normal, and the impact distance from the Solar neighborhood, $\xp$, during their penultimate, last and next disk crossings. We use these parameters to compute the direct response to the MW satellites and find that all of them excite bending modes and thus one-armed phase spirals in the Solar neighborhood, similar to that discovered in the Gaia data by \citet{Antoja.etal.18}. In the Solar vicinity, the largest direct response, by far, is due to the encounter with Sgr. The direct responses triggered by other satellites, most notably Hercules and the LMC, are at least $1-2$ orders of magnitude smaller. Hence, we conclude that, if the Gaia phase spiral was triggered by an encounter with a MW satellite, the strongest contender is Sgr. Although Sgr has been considered as the agent responsible for the Gaia phase spiral and other local asymmetries and corrugations, several studies have pointed out that it cannot be the sole cause of all these perturbations \citep[see e.g.,][]{Bennett.etal.21, Bennett.Bovy.21}. Our work argues, though, that the direct response in the Solar neighborhood from the other MW satellites, including the LMC, is not significant enough, at least in the range of actions covered by the Gaia snail. Of course, as discussed in section~\ref{sec::caveats}, the indirect response from the DM halo of the MW might play an important role especially for the more massive satellites such as Sgr and the LMC. Moreover the global response of a realistic disk will be different from that of the isothermal slab model considered here. We investigate the realistic disk response in Paper~II and leave a sophisticated analysis incorporating self-gravity and halo response for future work. It remains to be seen whether a combination of Sgr plus other (internal) perturbations due to for example spiral arms \citep[][]{Faure.etal.14} or the (buckling) bar \citep[e.g.,][]{Khoperskov.etal.19} can explain the fine-structure in the Solar neighborhood, or whether perhaps a solution requires modifying the detailed MW potential. It is imperative, though, to investigate the structure of phase spirals at other locations in the MW disk, in particular whether they are one-armed or two-armed. This would help to constrain both the time-scale and location of the perturbation responsible for the various out-of-equilibrium features uncovered in the disk of our MW.

\section*{Acknowledgments}

The authors are grateful to the anonymous referee for thoughtful comments and to Kathryn Johnston, Jason Hunt, Adrian Price-Whelan, Kaustav Mitra, Elena D'Onghia, Chris Hamilton and Dhruba Dutta-Chowdhury for insightful discussions and valuable suggestions. MW is supported by the National Science Foundation through Grant No. AST-1812689. FvdB is supported by the National Aeronautics and Space Administration through Grant No. 19-ATP19-0059 issued as part of the Astrophysics Theory Program.


\bibliography{references_banik}{}
\bibliographystyle{aasjournal}

\appendix

\section{Adiabatic limit of slab response}
\label{App:ad_lim_resp}

In the adiabatic/slow limit, the slab response can be computed by taking the $\omega_0\to 0$ limit and performing the $\tau$ integral in equation~(\ref{f1nk_isosol}) to obtain

\begin{align}
f_{1nk} = -i \pi\, \Phi_\rmN \calZ_n(I_z) \calX_k \left(\frac{n\Omega_z}{\sigma^2_z}+\frac{k v_x}{\sigma^2}\right) f_0(I_z,v_x,v_y)\, \delta(n\Omega_z+k v_x).
\end{align}
The Dirac delta function implies that only the resonant stars, i.e., those for which $n\Omega_z + k v_x=0$, contribute to the response in this slow limit. Substituting the expression for $f_0$ from equation~(\ref{f_iso}) in the above equation, integrating over $v_x$ and then summing over $n$, we obtain

\begin{align}
f_{1k} = -i\pi\,\Phi_\rmN \frac{\calX_k}{\left|k\right|} \sum_{n=-\infty}^{\infty} \calZ_n(I_z) \exp{\left[-\frac{n^2\Omega^2_z}{2 k^2\sigma^2}\right]} n\Omega_z \left(\frac{1}{\sigma^2_z}-\frac{1}{\sigma^2}\right) \exp{\left[i n w_z\right]}.
\end{align}
Substituting the Gaussian form for $\calX_k$ given in equation~(\ref{Phink_gaussian}) in the above expression, multiplying it by $\exp{\left[i k x\right]}$ and integrating over all $k$, we obtain the following final expression for the slab response in the slow limit:

\begin{align}
f_1(I_z,w_z,x) = -i \pi\,\Phi_\rmN \frac{\calX_k}{\left|k\right|} \sum_{n=-\infty}^{\infty} \calZ_n(I_z) \calJ_n(x)\, n\Omega_z \left(\frac{1}{\sigma^2_z}-\frac{1}{\sigma^2}\right) \exp{\left[i n w_z\right]},
\end{align}
where

\begin{align}
\calJ_n(x) = \int_{-\infty}^{\infty} d k\, \frac{\exp{\left[i k x\right]}}{\left|k\right|} \exp{\left[-k^2\Delta^2_x/2\right]} \exp{\left[-\frac{n^2\Omega^2_z}{2 k^2\sigma^2}\right]}.
\end{align}
The above integral can be approximately evaluated in the small and large $x$ limits by the saddle point method to obtain the following asymptotic behaviour of $\calJ_n(x)$:

\begin{align}
\calJ_n(x) &\sim 
\begin{cases}
\sqrt{\pi \sigma/2 \left|n\right| \Omega_z \Delta_x}\, \exp{\left[-\left|n\right|\Omega_z \Delta_x/\sigma\right]}\, \cos{\left(\sqrt{\frac{\left|n\right|\Omega_z}{\sigma \Delta_x}}x\right)}, & \text{small\;} x,\nonumber \\
\sqrt{2\pi}\,\frac{\Delta_x}{x} \exp{\left[-x^2/2\Delta^2_x\right]}, & \text{large\;} x.
\end{cases}
\end{align}
\\

\section{Slab response to satellite encounters}
\label{App:sat_disk_resp}

The perturbing potential, $\Phi_\rmP$, at $(x,z)$ due to a satellite galaxy impacting the disk along a straight orbit with uniform velocity $\vp$ at an angle $\thetap$ with respect to the normal is given by equation~(\ref{Phip_sat}). Computing the Fourier transform, $\Phi_{nk}$, of $\Phi_\rmP$, and substituting this in equation~(\ref{f1nk_isosol}) yields

\begin{align}
f_{1nk}(I_z,v_x,v_y,t) &=i\frac{G \Mp}{\vp} f_0(v_x,v_y,E_z) \left(\frac{n\Omega_z}{\sigma^2_z}+\frac{k v_x}{\sigma^2}\right) \exp{\left[-i\left(n\Omega_z+k v_x\right) t\right]}\, \calF_{nk}(t),
\label{f1nk_sat_1}
\end{align}
where

\begin{align}
\calF_{nk}(t) &= \frac{1}{{\left(2\pi\right)}^2} \int_0^{2\pi}d w'_z \exp{\left[-i n w'_z\right]} \int_{-\infty}^{\infty} d x' \exp{\left[-i k x'\right]} \int_{-\infty}^{t} d \tau\, \frac{\exp{\left[i\left(n\Omega_z+k v_x\right) \tau\right]}}{\sqrt{{\left(\tau-\frac{z'\cos{\thetap}+x'\sin{\thetap}}{\vp}\right)}^2+\frac{{\left(x'\cos{\thetap}-z'\sin{\thetap}\right)}^2}{\vp^2}}}.
\end{align}
The $\tau$ integral can be computed in the large $t$ limit to yield

\begin{align}
\calF_{nk}(t\to \infty) &= \frac{1}{2\pi^2} \int_0^{2\pi}d w'_z \exp{\left[-i n w'_z\right]} \int_{-\infty}^{\infty} d x' \exp{\left[-i k x'\right]} \nonumber \\
& \times \exp{\left[i \frac{\left(n\Omega_z+k v_x\right)\cos{\thetap} z'}{\vp} \right]} \exp{\left[i \frac{\left(n\Omega_z+k v_x\right)\sin{\thetap} x'}{\vp} \right]} K_0\left[\left(n\Omega_z+k v_x\right)\frac{\left(x'\cos{\thetap}-z'\sin{\thetap}\right)}{\vp}\right],
\end{align}
where $K_0$ denotes the zero-th order modified Bessel function of the second kind. Recalling that the unperturbed DF is isothermal, given by equation~(\ref{f_iso}), we integrate equation~(\ref{f1nk_sat_1}) over $v_x$ and $v_y$ to obtain

\begin{align}
&\int_{-\infty}^{\infty}d v_y\int_{-\infty}^{\infty}d v_x\, f_{1nk}(I_z,v_x,v_y,t) \approx \frac{\rho_c}{\sqrt{2\pi}\sigma_z} \exp{\left[-E_z/\sigma^2_z\right]} \frac{G\Mp}{\vp} \nonumber \\
&\times \frac{1}{2\pi^2} \int_0^{2\pi}d w'_z \exp{\left[-i n w'_z\right]} \exp{\left[i \frac{n\Omega_z\cos{\thetap} z'}{\vp} \right]} \int_{-\infty}^{\infty} d x' \exp{\left[-i k x'\right]} \exp{\left[i \frac{n\Omega_z\sin{\thetap} x'}{\vp} \right]} \nonumber \\
&\times \exp{\left[-\frac{1}{2}k^2\sigma^2 {\left(t-\frac{\calS}{\vp}\right)}^2\right]} \left[k^2 \left(t-\frac{\calS}{\vp}\right)+i\frac{n\Omega_z}{\sigma^2_z}\right] K_0\left[\left(n\Omega_z- i k^2 \sigma^2 \left(t-\calS/\vp\right)\right)\frac{\left(x'\cos{\thetap}-z'\sin{\thetap}\right)}{\vp}\right],
\label{f1nk_sat_2}
\end{align}
where we have defined

\begin{align}
\calS=z'\cos{\thetap}+x'\sin{\thetap}.
\end{align}
Multiplying equation~(\ref{f1nk_sat_2}) by $\exp{\left[ikx\right]}$ and integrating over $k$ yields

\begin{align}
&\int_{-\infty}^{\infty} d k\,\exp{\left[i k x\right]}\int_{-\infty}^{\infty}d v_y\int_{-\infty}^{\infty}d v_x\, f_{1nk}(I_z,v_x,v_y,t) \approx \frac{\rho_c}{\sqrt{2\pi}\sigma_z} \exp{\left[-E_z/\sigma^2_z\right]} \frac{G\Mp}{\vp} \nonumber \\ 
&\times \frac{1}{2\pi^2} \int_0^{2\pi}d w'_z \exp{\left[-i n w'_z\right]} \exp{\left[i \frac{n\Omega_z\cos{\thetap} z'}{\vp} \right]} \times \sqrt{2\pi} \int_{-\infty}^{\infty} d \Delta x\, \frac{1}{\sigma t'} \exp{\left[-\frac{1}{2}\frac{{(\Delta x)}^2}{\sigma^2 t'^2}\right]} \left[\frac{1}{\sigma^2 t'}\left(1+\frac{{(\Delta x)}^2}{\sigma^2 t'^2}\right)+i\frac{n\Omega_z}{\sigma^2_z}\right] \nonumber \\
&\times \exp{\left[i \frac{n\Omega_z\sin{\thetap} x'}{\vp} \right]} K_0\left[\left(n\Omega_z+ i \frac{{(\Delta x)}^2}{\sigma^2 t'^3}\right)\frac{\left(x'\cos{\thetap}-z'\sin{\thetap}\right)}{\vp}\right],
\end{align}
where $\Delta x = x-x'$, and $t'=t-\calS/\vp$. In the large time limit, using the identity that $\lim_{t'\to \infty} \exp{\left[-{(\Delta x)}^2/2\sigma^2 t'^2\right]}\Big/\sigma t'=\sqrt{2\pi}\delta (\Delta x)$, the integration over $\Delta x$ is simplified. Upon performing this integral, multiplying the result by $\exp{\left[i n w_z\right]}$ and summing over all $n$, we obtain the following response:

\begin{align}
f_1(I_z,w_z,x,t) &\approx \frac{\rho_c}{\sqrt{2\pi}\sigma_z} \exp{\left[-E_z/\sigma^2_z\right]}\times \frac{2G\Mp}{\vp} \nonumber \\
&\times \sum_{n=-\infty}^{\infty} \left[\frac{1}{\sigma^2 t}+i\frac{n\Omega_z}{\sigma^2_z}\right]\, \Psi_n(x,I_z)\, \exp{\left[i\,\frac{n\Omega_z \sin{\thetap}}{\vp}x\right]} \exp{\left[i n\left(w_z-\Omega_z t\right)\right]},
\label{f1_sat_app}
\end{align}
where
\begin{align}
\Psi_n(x,I_z)&= \frac{1}{2\pi} \int_0^{2\pi} d w_z\, \exp{\left[-i n \left(w_z - \frac{\Omega_z \cos{\thetap} z}{\vp}\right)\right]} K_0\left[\,\left|\frac{n\Omega_z \left(x\cos{\thetap}-z\sin{\thetap}\right)}{\vp}\right|\,\right].
\label{Psi_n_app}
\end{align}

The above expression for $\Psi_n$ can be simplified by evaluating the $w_z$ integral under the epicyclic approximation (small $I_z$ limit), to yield the following approximate form,
\begin{align}
\Psi_n(x,I_z) &\approx K_0\left(\frac{\left|n\Omega_z \cos{\thetap}\right|}{\vp}x\right) \Phi_n^{(0)}(I_z) - i\frac{n\Omega_z \sin{\thetap}}{\vp} K'_0\left(\frac{\left|n\Omega_z \cos{\thetap}\right|}{\vp}x\right) \Phi_n^{(1)}(I_z)\nonumber \\
&- \frac{1}{2} {\left(\frac{n\Omega_z \sin{\thetap}}{\vp}\right)}^2 K''_0\left(\frac{\left|n\Omega_z \cos{\thetap}\right|}{\vp}x\right) \Phi_n^{(2)}(I_z)+...\,.
\label{Psi_n_app_epi}
\end{align}
Here each prime denotes a derivative with respect to the argument of the function. $\Phi_n^{(j)}(I_z)$, for $j=0,1,2,...$, is given by
\begin{align}
\Phi_n^{(j)}(I_z) &= \frac{1}{2\pi} \int_0^{2\pi}d w_z\, z^j\, \exp{\left[-i n \left(w_z - \frac{\Omega_z \cos{\thetap} z}{\vp}\right)\right]} \nonumber \\
&\approx {\left(\frac{2I_z}{\nu}\right)}^{j/2} J_{n,j}\left(\frac{n\Omega_z \cos{\thetap}}{\vp}\sqrt{\frac{2I_z}{\nu}}\right).
\label{Phin_sat_app}
\end{align}
Here the implicit relation between $z$, $w_z$ and $I_z$ given in equation~(\ref{z_wz_Iz}), which yields $z=\sqrt{2 I_z/\nu}\, \sin{w_z}$ for small $I_z$, has been used. $J_{n,j}$ denotes the $j^{\rm th}$ derivative of the $n^{\rm th}$ order Bessel function of the first kind, and $\nu=\sqrt{2}\,\sigma_z/h_z$ is the vertical epicyclic frequency. In equation~(\ref{f1_sat_app}), well after the encounter (large $t$), the term, $1/\sigma^2 t$, can be neglected relative to $i n\Omega_z/\sigma^2_z$ for $n\neq 0$, thus yielding the expression for the disk response to satellite encounters given in equation~(\ref{f1_sat}).

\section{Detectability criterion for the phase spiral}
\label{App:detect_crit}

The demarcation between strong and weak amplitudes of a phase spiral is dictated by the minimum detectable relative response, $\delta$, which can be determined in the following way. Let there be a phase spiral that we want to detect with a total number, $N_*$, of stars by binning the phase-space distribution in the $\sqrt{I_z}\cos{w_z}-\sqrt{I_z}\sin{w_z}$ plane. Let us define the unperturbed DF, $f_0$, and the normalized unperturbed DF, $\bar{f}_0$, such that

\begin{align}
N_* = \iint f_0\, d I_z\, d w_z,\;\;\;\; \bar{f}_0 = \frac{f_0}{N_*}.
\end{align}
The perturber introduces a perturbation in the (normalized) DF, $\bar{f}_1$, which manifests as a spiral feature in the phase-space due to phase-mixing. To recover $\bar{f}_1$ we bin the data in $I_z$ and $w_z$, such that the perturbation in the number of stars in each bin ($\Delta I_z,\Delta w_z$) is given by

\begin{align}
N(\Delta I_z,\Delta w_z) = N_* \bar{f}_1 \Delta I_z \Delta w_z.
\end{align}
The optimum binning strategy can be determined as follows. The phase spiral is a periodic feature in both $I_z$ and $w_z$. Therefore, to pull out the periodicity in $I_z$, we need to sample with a frequency exceeding the Nyquist frequency, i.e., the bin size, $\Delta I_z$, should be less than $I_{z,\rm max}/N_{\rm wind}$, where $I_{z,\rm max}$ is the maximum $I_z$ in the sample and $N_{\rm wind}$ is the number of winds of the spiral. Moreover, $\Delta I_z$ is required to exceed the Gaia measurement error so that the error is dominated by Poisson noise, i.e., we require $\Delta I_z/I_z > \Delta_{\rm Gaia} \sim 10^{-2}$ \citep[see][for parallax and radial velocity errors, the two dominant sources of measurement errors in Gaia]{Luri.etal.18,Katz.etal.19}. Within each $I_z$ bin, the data is further divided into $N_a$ azimuthal bins, each of size $\Delta w_z=2\pi/N_a$. For optimum sampling in $w_z$, $N_a$ should be greater than $2n$ (for spiral mode $n$) and less than $2\pi/\Delta_{\rm Gaia}$. After binning the data as discussed above, a reliable detection of the phase spiral can be made with a given signal to noise ratio, $S/N$, when the perturbation in the number of stars in each bin,

\begin{align}
N(\Delta I_z,\Delta w_z) = N_* \times \frac{\bar{f}_1}{\bar{f}_0} \times \frac{2\pi \bar{f}_0(I_z) \Delta I_z}{N_a} \geq {\left(S/N\right)}^2.
\end{align}
Here we have assumed that the error in recovering the spiral feature is dominated by Poisson noise. This yields the following estimate for the minimum detectable relative response for an isothermal slab,

\begin{align}
\frac{\bar{f}_1}{\bar{f}_0} \geq \delta = 3.6\times 10^{-4} \times {\left(\frac{S/N}{3}\right)}^2 \left(\frac{10^6}{N_*}\right) \left(\frac{N_a}{10}\right) \left(\frac{0.1}{\Delta I_z/I_z}\right) \frac{h_z\sigma_z}{I_z}\, \exp{\left[\frac{E_z(I_z)}{\sigma^2_z}\right]}.
\end{align}
Provided that there are about a million stars in the Gaia data of the Solar neighborhood \citep[][]{Antoja.etal.18}, we consider $\delta=10^{-4}$ to be a rough estimate for the minimum detectable relative response.

\label{lastpage}

\end{document}